\documentclass[pdflatex,sn-mathphys-num]{sn-jnl}

\usepackage{graphicx}%
\usepackage{multirow}%
\usepackage{amsmath,amssymb,amsfonts}%
\usepackage{amsthm}%
\usepackage{mathrsfs}%
\usepackage[title]{appendix}%
\usepackage{xcolor}%
\usepackage{textcomp}%
\usepackage{manyfoot}%
\usepackage{booktabs}%
\usepackage{algorithm}%
\usepackage{algorithmicx}%
\usepackage{algpseudocode}%
\usepackage{listings}%

\theoremstyle{thmstyleone}%
\theoremstyle{thmstyletwo}%

\theoremstyle{thmstylethree}%

\usepackage{bm}
\usepackage{here}
\usepackage{graphicx}
\usepackage{tikz}
\usepackage{xcolor}
\usepackage{circuitikz}
 \usepackage{pdflscape}
 \pgfcircdeclarebipolescaled{instruments}
{
    \anchor{text}{\pgfextracty{\pgf@circ@res@up}{\northeast}
        \pgfpoint{-.5\wd\pgfnodeparttextbox}{
            \dimexpr.5\dp\pgfnodeparttextbox+.5\ht\pgfnodeparttextbox+\pgf@circ@res@up\relax
        }
    }
}
{\ctikzvalof{bipoles/oscope/height}}
{josephson}
{\ctikzvalof{bipoles/oscope/height}}
{\ctikzvalof{bipoles/oscope/width}}
{
    \pgf@circ@setlinewidth{bipoles}{\pgfstartlinewidth}
    \pgfextracty{\pgf@circ@res@up}{\northeast}
    \pgfextractx{\pgf@circ@res@right}{\northeast}
    \pgfextractx{\pgf@circ@res@left}{\southwest}
    \pgfextracty{\pgf@circ@res@down}{\southwest}
    \pgfmathsetlength{\pgf@circ@res@step}{0.25*\pgf@circ@res@up}
    \pgfscope
        \pgfpathrectanglecorners{\pgfpoint{\pgf@circ@res@left}{\pgf@circ@res@down}}{\pgfpoint{\pgf@circ@res@right}{\pgf@circ@res@up}}
        \pgf@circ@draworfill
    \endpgfscope
    \pgfscope
      \pgfpathmoveto{\pgfpoint{\pgf@circ@res@left}{\pgf@circ@res@up}}%
      \pgfpathlineto{\pgfpoint{\pgf@circ@res@right}{\pgf@circ@res@down}}%
      \pgfpathmoveto{\pgfpoint{\pgf@circ@res@right}{\pgf@circ@res@up}}%
      \pgfpathlineto{\pgfpoint{\pgf@circ@res@left}{\pgf@circ@res@down}}%
      \pgfusepath{draw}
    \endpgfscope
}
\def\pgf@circ@josephson@path#1{\pgf@circ@bipole@path{josephson}{#1}}
\tikzset{josephson/.style = {\circuitikzbasekey, /tikz/to path=\pgf@circ@josephson@path, l=#1}}

\begin{document}

\title[Ohm's law, Joule heat, and Planckian dissipation]{Ohm's law, Joule heat, and Planckian dissipation}


\author*[ ]{\fnm{Hiroyasu} \sur{Koizumi}}\email{koizumi.hiroyasu.fn@u.tsukuba.ac.jp}



\affil*[ ]{\orgdiv{Center for computational sciences}, \orgname{University of Tsukuba}, \orgaddress{\street{1-1-1 Tennodai}, \city{Tsukuba}, \postcode{305-8577}, \state{Ibaraki}, \country{Japan}}}





\abstract{Electric current generation and its dissipation are important physical processes.
It ranges from the one follows the Ohm's law to superconductivity.
Recently, it has been shown that the gradient of the chemical potential force arises from the time-component of the Berry connection from many-electron wave functions, and we consider its importance for the electric current conduction in this work.
We first show that it rectifies the odd explanation in Joule heating 
by electric current in a metallic wire: Poynting's theorem explains that the energy for the Joule heating enters from the outside of the wire as radiation. 
We show that this energy is supplied by the chemical potential gradient generated by the battery connection. 
Next, we consider the discharging of a capacitor problem where the capacitor plays a role of a battery; and the tunneling supercurrent through the Josephson junction problem, where the original derivation did not include the capacitor contribution.
Lastly, we argue that
the gauge fluctuation of the time-component of the Berry connection included in the chemical potential gradient force might
explain the Planckian dissipation observed in high transition temperature cuprate superconductors. The present work suggests the rethinking of the gauge invariance in Maxwell's equations.}




\maketitle

\section{Introduction}

One of old and established physical laws for the electric current is Ohm's law
discovered by Ohm \cite{Ohm} (unpublished materials indicate it was known before Ohm by Cavendish \cite{cavendish}): The electric current $I$ generated by connecting a battery of voltage $V$ to a metallic wire with resistance $R$ is given by
\begin{eqnarray}
I={V \over R}
\label{Ohm}
\end{eqnarray}
The battery originally used by Ohm was an electric battery, where the electromotive force was generated by chemical reactions.

Joule found that the heat generated by a metallic wire of resistance $R$ with current $I$ is given by
\begin{eqnarray}
RI^2=IV
\label{Joule}
\end{eqnarray}
where the Ohm's law is used \cite{Joule}. It indicates the heat is equivalent to the work done by the battery, establishing that the heat is a kind of energy. 

Surprisingly, this old, thought to be established,  dissipation phenomenon has loose ends that should be tied: The energy flow explained by the Poynting theorem for the above Joule heat generation is known to be rather odd \cite{FeynmanII27-5,Poynting2022}.

Let us first explain the Poynting theorem.
The conservation of the energy in systems composed of electromagnetic field and electric current was formulated by
Poynting and called, `Poynting's theorem' \cite{Poynting}. It states that
the rate of the energy consumed by the current in the volume ${\cal V}$ is given by $\int_{\cal V} d^3r \ {\bf j} \cdot  {\bf E}$, where ${\bf j}$ and ${\bf E}$ are the current density and electric field, respectively;
and the rate of the energy leaving ${\cal V}$  as radiation is given by $\int_{\cal S} d{\bf S} \cdot ({\bf E} \times {\bf H})$, where ${\bf B}$ is the magnetic field, ${\bf H}$ is related to ${\bf B}$ by ${\bf H}=\mu_0^{-1} {\bf B}$ ($\mu_0$ is the vacuum permeability) in a vacuum, and ${\cal S}$ is the boundary surface of the volume ${\cal V}$.
In short, the energy conservation is given by
\begin{eqnarray}
\int_{\cal V} d^3r \  {\bf j}  \cdot  {\bf E}+
\int_{\cal S} d{\bf S} \cdot ({\bf E} \times {\bf H})=-{ \partial \over {\partial t}}\int_{\cal V} d^3r \  u
\label{eqPoynting}
\end{eqnarray}
where $u$ is the energy density of the electromagnetic field. The first term looks similar to Eq.~(\ref{Joule}); thus,
 it is often said that it expresses the Joule heat.

The Ohm's law was explained by Drude \cite{Drude}, and later by Lorentz \cite{Lorentz1905,Lorentz} and Bohr \cite{Bohr1911}, assuming that the battery connected to the wire generates the electric field ${\bf E}_{\rm Drude}$ inside it. 
The law is expressed in a local field relation, 
\begin{eqnarray}
{\bf j}= \sigma {\bf E}_{\rm Drude}
\label{eqOhm}
\end{eqnarray}
where $\sigma$ is the conductivity of the wire.
Then, ${\bf j}  \cdot  {\bf E}$ in the first term in Eq.~(\ref{eqPoynting}) is given by
 ${\bf j} \cdot {\bf E}_{\rm Drude} =\sigma^{-1} {\bf j}^2 > 0$; here, the plus sign indicates that it is the energy consumed.
 
The oddness of the energy flow is that the radiation enters into the wire from outside as expressed by the Poynting vector ${\bf E} \times {\bf H}$ in Eq.~(\ref{eqPoynting}), and supplies the energy for the Joule heat consumed in the wire.
However, it is sensible to consider that the energy for the Joule heat is supplied by the battery through the current flows in the wire.

 This problem was taken up by the present author, recently \cite{Koizumi2024Lorentz}. A sensible explanation is obtained by identifying that the gradient of the chemical potential force is the one accelerates electrons instead of the electric field. 
The chemical potential was originally introduced by Gibbs (he called it, the intrinsic potential) in the context of thermodynamics 
\cite{Gibbs}.  During the time when Drude developed his theory, the chemical potential was unknown; however, we now know the chemical potential exists, and the electromotive force generated is due to the
chemical potential difference originates from chemical reactions in the battery.
The chemical potential appears as the Fermi energy in solid state physics, and its gradient is known to produce current.
 It is also used in the Landauer-B\"{u}ttiker theory \cite{Landauer,Buttiker1985,Datta}, where 
the chemical potential appears as the Fermi energy of the distribution function of the free-electron theory of conduction electrons. 

It has been shown recently that this chemical potential arises as a quantum many-body effect described by the Berry connection \cite{Koizumi2024Lorentz}. This indicates the appearance of the Berry connection and also of the chemical potential are general quantities of quantum many-body systems \cite{koizumi2022,koizumi2023} that include $\hbar$ ($\hbar$ is the reduced Planck constant).

Recently, the Planckian dissipation observed in strange metals such as high transition temperature cuprate superconductors  is a focus of attention in condensed matter physics \cite{Sachev1992,Zaanen2019,Planckian}. It is characterized by the
relaxation time with $\hbar$, given by $\tau_{\hbar}= {\hbar \over {k_B T}}$ ($k_B$ is the Boltzmann constant, and $T$ is temperature).
The elucidation of the origin of it is considered to be the key to understand
high transition temperature cuprate superconductivity. One of the purposes of the present work is 
to show that this Planckian dissipation may be related to the fluctuation of the Berry connection that produces the chemical potential.
If this is the case, the loose ends of the Joule heating problem
and the Planckian dissipation in the cuprate superconductivity have the common origin. 

The Berry connection from many-electron wave functions was originally put forward by the present author
to develop a new superconductivity theory \cite{koizumi2022,koizumi2023} that encompasses the standard theory based on the BCS one \cite{BCS1957}. The initial motivation for the development of it is to elucidate the
mechanism of the cuprate superconductivity; however, during such an effort, it became apparent that experimental facts of superconductivity indicate the presence of several serious loose ends to be tied in the standard theory (we will discuss them in Section~\ref{Section3}). 
The new theory was made to tie them with keeping major successful results of the BCS theory intact.
This is achieved by introducing new 
Bogoliubov type excitations; they preserve the particle number unlike the original ones, and the Nambu-Goldstone like mode arises form the Berry connection with keeping the particle number fixed (A short account of this theory and its relation to the BCS theory is given in \ref{Bogoliubov}).

In the new theory, the Berry connection produces persistent current if it is equipped with nontrivial topological quantum numbers. 
If the fluctuation of the topological quantum numbers for the Berry connection (we will call it, the `gauge fluctuation') occurs, the current becomes resistive.  Such a resistive current is expected to exist in the vicinity of a superconducting phase of the phase diagram, and may exhibit the Planckian dissipation. We will consider this possibility in the present work.

Before dealing with the Planckian dissipation problem, we will first revisit the
Joule heating problem and show how the standard textbook explanation is rectified to a sensible explanation in which the energy is supplied from the battery through the current.
We also consider the discharging of a capacitor; then, the supercurrent flow through the Josephson junction including the capacitor contribution. 
It is noteworthy that although the Josephson junction inherently has a capacitor contribution due to its structure, it was not included in the original Josephson's work \cite{Josephson62} or the textbook explanation \cite{FeynmanIII21-9}.
The derivation including the capacitance contribution indicates that the electrons transfer singly across the junction, in contrast to the Josephson's derivation
that assumes the pairwise transfer \cite{Josephson62}. Note that this does not mean the absence of the stabilization by the electron-pairing; the experimentally observed flux quantum is still obtained.

Actually, the existence of single electrons in superconductors has been speculated for a long time to explain the Knight shift experiments \cite{SuperBook8-8}. It is also noteworthy that a recent experiment indicates the absence of the dissipative quantum phase transition in a Josephson junction system predicted by the standard theory \cite{PhysRevX2021a};
and also the so-called `quasiparticle poisoning problem' indicates the existence of a large amount of excited single electrons in Josephson junction systems for qubits, obtaining the observed ratio of their number to the Cooper pair number  $10^{-9} \sim 10^{-5}$ in disagreement with the standard theory ratio $10^{-52}$ \cite{poisoning2023,Serniak2019}.
 
The organization of the present work is as follows: First, we briefly summarize the Berry connection from many-electron wave functions, and explain the appearance of the gradient of the chemical potential force from it. Next, we revisit the plasma oscillation and screening in the electron gas model including the
gradient of the chemical potential force; it is indicated the force balance exists for
the force from the static electric field and that from the chemical potential gradient. A similar force balance is assumed in the Ohm's law situation, later.
Then,  we will proceed to put forward a new criterion to differentiate the normal and superconducting states based on the stability of the velocity field with the Berry connection equipped with non-trivial topological quantum numbers.

 Next, we consider the energy flow for electric conduction in a metallic wire connected to a battery; we argue that the battery connection generates 
the gradient of the chemical potential inside the wire, and this is the force accelerates conduction electrons; when the steady current flow is established, the force balance exists between
the gradient of the chemical potential force, and that from the induced electric field.
Then, we consider the energy flow during the discharging of a capacitor through a resistive wire; in this case, the capacitor acts as the battery for the current generation. 
Next, we discuss the Josephson effect in superconductors with including the capacitor contribution.
Then, the Planckian dissipation is examined by considering the gauge fluctuation of the Berry connection. Lastly, we conclude the present work by discussing its implications in the interpretation of the Maxwell's equations.

\section{Forces on electrons in an electromagnetic field including the Berry connection from many-electron wave functions}

Let us consider the Berry connection  from many-body wave functions defined by
\begin{eqnarray}
\!{\bf A}^{\rm MB}_{\Psi}({\bf r},t)\!=\!
{{{\rm Re} \left\{
 \int d\sigma_1  d{\bf x}_{2}  \cdots d{\bf x}_{N}
 \Psi^{\ast}({\bf r}, \sigma_1, \cdots, {\bf x}_{N},t)
  (-i \hbar \nabla )
\Psi({\bf r}, \sigma_1, \cdots, {\bf x}_{N},t) \right\}
 }
 \over {\hbar  N^{-1}}\rho({\bf r},t)}
 \nonumber
 \\
 \label{Berry}
\end{eqnarray}
Here,
`$\rm{Re}$' denotes the real part, $\Psi$ is the total wave function, ${\bf x}_i=({\bf r}_i, \sigma_i)$ collectively stands for the coordinate ${\bf r}_i$ and the spin $\sigma_i$ of the $i$th electron, however, ${\bf x}_1$ is expressed as  ${\bf x}_1=({\bf r}, \sigma_1)$; $N$ is the total number of electrons; $-i \hbar \nabla$ is the Schr\"{o}dinger's momentum operator for the coordinate vector ${\bf r}$, and $\rho$ is the number density calculated from $\Psi$ \cite{Koizumi2024Lorentz,koizumi2023}.

From the original wave function $\Psi$, we can construct the following currentless wave function using the Berry connection ${\bf A}_{\Psi}^{\rm MB}$
\begin{eqnarray}
\Psi_0 ({\bf x}_1, \cdots, {\bf x}_{N},t)=\Psi ({\bf x}_1, \cdots, {\bf x}_{N},t)\exp\left(-i \sum_{j=1}^{N} \int_{0}^{{\bf r}_j} {\bf A}_{\Psi}^{\rm MB}({\bf r}',t) \cdot d{\bf r}' \right)
\nonumber
\\
\label{wavef-1}
\end{eqnarray}
The current density calculated by $\Psi_0$ is zero due to the cancellation of the current density from $\Psi$ and that
from $\exp\left(-i \sum_{j=1}^{N} \int_{0}^{{\bf r}_j} {\bf A}_{\Psi}^{\rm MB}({\bf r}',t) \cdot d{\bf r}' \right)$.
Conversely, $\Psi$ is expressed using $\Psi_0$ as 
\begin{eqnarray}
\Psi ({\bf x}_1, \cdots, {\bf x}_{N},t)=\exp\left(i \sum_{j=1}^{N} \int_{0}^{{\bf r}_j} {\bf A}_{\Psi}^{\rm MB}({\bf r}',t) \cdot d{\bf r}' \right)\Psi_0 ({\bf x}_1, \cdots, {\bf x}_{N},t)
\nonumber
\\
\label{wavef0}
\end{eqnarray}
where the electric current generation is entirely attributed to the factor
\begin{eqnarray}
 \exp ~({ i \sum_j \int_{0}^{{\bf r}_j} {\bf A}_{\Psi}^{\rm MB}({\bf r}',t) \cdot d{\bf r}'} )
 \end{eqnarray}
 This acts as the collective motion of electrons that generates electric current.

The velocity field ${\bf v}$ for the electrons is calculated using $\Psi$  with including the electromagnetic vector potential ${\bf A}$ as
 \begin{eqnarray}
{\bf v}={e \over m_e}{\bf A}+{\hbar \over {m_e}}{\bf A}_{\Psi}^{\rm MB}
\label{eq12v2-0}
\end{eqnarray}
where $-e$ is the electron charge and $m_e$ is the electron free mass.
Up to this point, no approximation is made.

Now, we introduce angular variable $\chi$ with period $2\pi$ and assume that it is related to ${\bf A}_{\Psi}^{\rm MB}$ 
in the following manner 
 \begin{eqnarray}
{ {\chi({\bf r},t)}}= - 2\int^{{\bf r}}_0 {\bf A}_{\Psi}^{\rm MB}({\bf r}',t) \cdot d{\bf r}' 
\label{int-chi}
\end{eqnarray}
The reason for the introduction of this $\chi$ will be explained, later (see also \ref{Bogoliubov}, around Eq.~(\ref{eqA9})).

Then, $\Psi$ is expressed as
\begin{eqnarray}
\Psi ({\bf x}_1, \cdots, {\bf x}_{N},t)=\exp\left(-{i \over 2}\sum_{j=1}^{N}\chi({\bf r}_j,t)\right)
\Psi_0 ({\bf x}_1, \cdots, {\bf x}_{N},t)
\nonumber
\\
\label{wavef1}
\end{eqnarray}
and ${\bf v}$ becomes
 \begin{eqnarray}
{\bf v}
={e \over m_e}\left({\bf A}-{\hbar \over {2e}}\nabla \chi \right)
\label{eq12v2}
\end{eqnarray}

Next, we consider the sum, ${\bf A}-{\hbar \over {2e}}\nabla \chi$, in ${\bf v}$. It is gauge invariant due to the fact that the gauge ambiguity of ${\bf A}$ is compensated by the ambiguity of $\nabla \chi$; in other words, the value of ${\bf A}-{\hbar \over {2e}}\nabla \chi$ is the same irrespective of the gauge choice made for ${\bf A}$ 
since during the evaluation process of $\nabla \chi$, this freedom is absorbed in the evaluated $\nabla \chi$ to give the same value for ${\bf A}-{\hbar \over {2e}}\nabla \chi$ (this point is shown in the calculations performed in \cite{Koizumi2021c}). 

The gauge invariance of ${\bf A}-{\hbar \over {2e}}\nabla \chi$ implies its time-component partner,
${\varphi}+{\hbar \over {2e}}\partial_t \chi $, 
is also gauge invariant. Note that the electromagnetic four-vector gauge potential $({\bf A}, \varphi)$ becomes $({\bf A}-{\hbar \over {2e}}\nabla \chi, \varphi+{\hbar \over {2e}}\partial_t \chi )$ with including the Berry connection.
Taking into account the fact that the chemical potential $\mu$ appears in a similar manner as the 
scalar potential $\varphi$ in the electron Hamiltonian, and that the chemical potential is gauge invariant (however, $\varphi$ is not), we obtain the following relation
\begin{eqnarray}
\mu =e\left(
\varphi + { \hbar \over {2e}}\partial_t \chi
\right)
\label{eq1mu}
\end{eqnarray}
This is the chemical potential we use in this work (this relation is also obtained as Eq.~(\ref{H1}) in \ref{Bogoliubov} from
a different reasoning). Actually, a similar relation is obtained by Anderson for the case of superfluid helium \cite{Anderson66}.

Let us calculate the force acting on electrons by $m_e{{d{\bf v}} \over {dt}}$,
\begin{eqnarray}
m_e{{d{\bf v}} \over {dt}}
&=&m_e \left[\partial_t{\bf v}+
{1 \over 2}\nabla v^2-{\bf v}\times(\nabla \times {\bf v}) \right]
\nonumber
\\
&=&e  \partial_t {\bf A}-{ \hbar \over {2}}\partial_t(\nabla \chi)
+{m_e \over 2}\nabla v^2-{ e }{\bf v}\times{\bf B}
+{ \hbar \over {2}}{\bf v}\times \nabla \times (\nabla \chi)
\nonumber
\\
&=&{ {-e}}{\bf E}-\nabla \mu
+{m_e \over 2}\nabla v^2-e{\bf v}\times{\bf B}
\nonumber
\\
&&
+{ \hbar \over {2}}{\bf v}\times \nabla \times (\nabla \chi)
+{ \hbar \over {2}}(\nabla \partial_t -\partial_t \nabla )\chi
\label{eq13a}
\end{eqnarray}
where Eqs.~(\ref{eq12v2}) and (\ref{eq1mu}), and the following relations, ${\bf B}=\nabla \times {\bf A}$,  ${\bf E}=-\partial_t {\bf A} -\nabla \varphi$  are used. 
The forces in Eq.~(\ref{eq13a}) include the standard Lorenz force ${ {-e}}{\bf E}$$-e{\bf v}\times{\bf B}$, and additional forces.
The important one in this work is the gradient of the chemical potential force, $-\nabla \mu$.

At this point, it is worth mentioning the argument given by Maxwell in his book on electromagnetism \cite{Maxwell-297}: The electromotive intensity at any point is the resultant force on a unit of positive electricity placed at that point. It may arise from the following three actions

\begin{enumerate}

\item[(1)] Electrostatic action

\item[(2)] Electromagnetic induction

\item[(3)] Thermoelectric or electrochemical action

\end{enumerate}
The gradient of the chemical potential force belong to the third category.
The first two are included in the standard textbook, but the third one is often not mentioned.
There are terms in Eq.~(\ref{eq13a}) that are not categorized above. They are important in superconductivity
as shown in our previous work \cite{Koizumi2024Lorentz}.

In superconductors, ${\bf v}$ in Eq.~(\ref{eq12v2}) can be nonzero, and stable.
Then, the current density is given by
\begin{eqnarray}
{\bf j}=-en_s{\bf v}=-{{e^2 n_s} \over m_e}\left({\bf A}-{\hbar \over {2e}}\nabla \chi \right)
\label{eq2J}
\end{eqnarray}
where $n_s$ is the number density of electrons that flow with ${\bf v}$; we explain the 
meaning of $n_s$, below. 

In the wave function in Eq.~(\ref{wavef1}), all $N$ electrons participate in the `$\chi$ mode' described by the factor $e^{-{i \over 2}\chi}$. However, it is possible that the number can be different from $N$. In other words, the wave function is given as a superposition of states with different number of electrons in the $\chi$ mode,
\begin{eqnarray}
\Psi({\bf x}_1, \cdots, {\bf x}_N,t)&=&C_0 \Psi_{00}({\bf x}_1,  \cdots, {\bf x}_N,t)
\nonumber
\\
&+&
C_1 \sum_{j_1=1}^N e^{  -{i \over 2} \chi({\bf r}_{j_1},t)}\Psi_{01}({\bf x}_1,  \cdots, {\bf x}_N,t)
\nonumber
\\
&+&C_2 \sum_{j_1 > j_2} e^{-{i \over 2} [\chi({\bf r}_{j_1},t) + \chi({\bf r}_{j_2},t) ]}\Psi_{02}({\bf x}_1,  \cdots, {\bf x}_N,t)
\nonumber
\\
&+&C_3 \sum_{j_1 > j_2 > j_3} e^{  -{i \over 2} [\chi({\bf r}_{j_1},t) + \chi({\bf r}_{j_2},t) 
+ \chi({\bf r}_{j_3},t)]}\Psi_{03}({\bf x}_1,  \cdots, {\bf x}_N,t)
\nonumber
\\
&&\cdots \cdots
\nonumber
\\
&+&C_N e^{ -{i \over 2} [ \chi({\bf r}_1,t)+ \cdots + \chi({\bf r}_N,t) ]}
\Psi_{0N}({\bf x}_1,  \cdots, {\bf x}_N,t)
\nonumber
\\
\label{single-valued3}
\end{eqnarray}
where $C_0, \cdots, C_N$ are constants; $\Psi_{00}, \cdots, \Psi_{0N}$ are currentless antisymmetric wave functions. Note that $\Psi$ is antisymmetric with respect to the exchange of electrons.
$\Psi$ in Eq.~(\ref{wavef1}) corresponds to the case $C_0=C_1=\cdots=C_{N-1}$, $C_N=1$ and $\Psi_{0N}=\Psi_{0}$. 

Using this fluctuation of the number of electrons participating in the $\chi$ mode,
the system can reduce the total energy.
A practical way to include this fluctuation is to use the modified Bogoliubov-de Gennes type equation,
which is explained succinctly in \ref{Bogoliubov}. The important ingredients in this formalism 
are the particle number conserving Bogoliubov operators $\gamma_n$ and $\gamma_n^\dagger$ that describe the transitions between states with different numbers of 
electrons in the $\chi$ mode. The superconducting electron density $n_s$ is given as the number density of electrons participating in the $\chi$ mode, 
$\rho_\chi$, in \ref{Bogoliubov}.

From the viewpoint of the new theory, the BCS theory is the theory that takes into account the number fluctuation of electrons in the $\chi$ mode as the number fluctuation of Cooper pairs; the convenient formalism to include such fluctuations is to use
the particle number nonconserving Bogoliubov operators \cite{Bogoliubov58,Valatin}. In this respect, the BCS theory is a theory using the gauge symmetry approximation \cite{Peierls1991,Peierls92}.
The normal ground state solution corresponds to the case where $C_0=1$, $C_1=\cdots=C_{N-1}=C_N=0$ as explained by Bloch and Bohm \cite{Bohm-Bloch}, which yields $n_s=0$. 

Let us examine the supercurrent generated by ${\bf v}$. Since ${\bf v}$ is gauge invariant, we may choose ${\bf A}$ in the gauge $\nabla \cdot {\bf A}=0$.
Then, the conservation of the charge for the stationary case, $\nabla \cdot {\bf j}=0$,  yields
\begin{eqnarray}
\nabla^2 \chi =0
\label{eq2Chi}
\end{eqnarray}
where the spatial dependence of $n_s$ is neglected for simplicity.
The solution to the above equation is characterized by an integer called the `winding number',
\begin{eqnarray}
w[\chi]_C={1 \over {2\pi}}\oint_C \nabla \chi \cdot d{\bf r}
\label{eq2W}
\end{eqnarray}
where $C$ is a loop in the space. If this number is non-zero, a persistent loop current flows.
If a collection of such loop currents is stable, a macroscopic persistent current, `supercurrent' will be realized.
This supercurrent yields the flux quantization in the unit of ${h \over {2e}}$ as observed in superconductors.

In the new theory, the appearance of  $\chi$ in Eq.~(\ref{int-chi}) is attributed to the following 
two constrains on the solution of the Schr\"{o}dinger equation: 

\begin{enumerate}

\item The conservation of local charge. 

\item The single-valuedness of the wave function as a function of the electron coordinates. 

\end{enumerate}

For the cuprate superconductivity case, the present author has been arguing that the
spinor property of electrons plays an important role.
It causes sign-change of the spin-function when spin is twisted one around
for the excursion along the loop; then, in order to fulfill the second constraint,
$\chi$ appears on the coordinate function to compensate the spin function sign change. 
It can be shown that if we combine the above two constrains, we can obtain$\nabla \chi$,
which yields $\chi$ with non-zero winding numbers \cite{koizumi2023,Koizumi2021c}.

It is noteworthy that the collective plasma mode is obtained in the electron gas model by a constraint on the electron density, which is similar to the first constraint above \cite{BohmPines}. 
In the new theory, the second one is added; it is needed to calculate nonzero current density using the wave function.

 Anderson found the new collective mode in the BCS model due to the presence of Cooper pairs 
\cite{Anderson1958a,Anderson1958b}. In the new theory, this mode is replaced by the $\chi$ mode.
As the wave function in Eq.~(\ref{single-valued3}) indicates, it allows the single particle change of the number of electrons participating in it. This enables the Josephson tunneling by single-electron.

\section{Plasma oscillation and screening}
\label{Plasma}

Extending the theoretical method that yields the plasma mode in the presence of Cooper pairs,
 Anderson found the new collective mode in the BCS model 
\cite{Anderson1958a,Anderson1958b}. This mode corresponds to the $\chi$ mode in the new theory. In this section, we examine the effect of $\chi$ mode in the plasma oscillation and screening. The effect we consider is that arising from 
the gradient of the chemical potential force. The non-zero winding number for $\chi$ does not play any role here, although it is important in superconductivity.

We follow the Fetter and Walecka's book \cite {Fetter-Walecka} with adding the gradient of the chemical potential force:
Let us consider an electron gas with electron density $\rho$ and the background compensating positive charge density $\rho_0$. 
If the electron density is slightly perturbed from the equilibrium value $\rho_0$
\begin{eqnarray}
\rho({\bf r}, t)=\rho_0 + \delta \rho({\bf r}, t)
\end{eqnarray}
an electric field ${\bf E}$ that satisfies
\begin{eqnarray}
\epsilon_0 \nabla \cdot {\bf E}= - e \delta \rho({\bf r}, t)
\label{eq20}
\end{eqnarray}
arise.

The forces acting on the electrons are the one from the electric field and another from the gradient of the chemical potential. Thus, the Newtonian equation yields
\begin{eqnarray}
m_e{ d \over {dt}}(\rho {\bf v})=- e\rho{\bf E} -\rho \nabla \mu 
\end{eqnarray}
or 
\begin{eqnarray}
m_e \rho_0 { \partial \over {\partial t}}{\bf v} \approx -e\rho_0{\bf E} -\rho_0 \nabla \mu 
\label{eq22}
\end{eqnarray}
Another relation is obtained from the equation of continuity,
\begin{eqnarray}
 { {\partial \rho}  \over {\partial t}}+ \nabla \cdot (\rho {\bf v})\approx 
 { {\partial \delta \rho}  \over {\partial t}}+ \rho_0\nabla {\bf v}
 =0
 \label{eq23}
\end{eqnarray}

From Eqs.~(\ref{eq20}), (\ref{eq22}), and (\ref{eq23}), we obtain
\begin{eqnarray}
 { {\partial^2 \delta \rho}  \over {\partial t^2}}&=& -\rho_0 {\partial \over {\partial t}} \nabla \cdot {\bf v}
 \nonumber
 \\
 &=&{{e \rho_0} \over m_e}(\nabla \cdot {\bf E}+e^{-1}\nabla^2 \mu)
 \nonumber
 \\
 &=&-\Omega_{pl}^2\delta \rho+{\rho_0 \over m_e}\nabla^2 \mu
 \label{eq24}
 \end{eqnarray}
where the plasma frequency is defined by 
\begin{eqnarray}
\Omega_{pl}=\left( {{e^2 \rho_0} \over {\epsilon_0 m_e}}\right)^{1/2}
 \end{eqnarray}
 If the gradient of the chemical potential force is neglected, the oscillation occurs with the frequency $\Omega_{pl}$. 
 
 Let us consider the effect of the gradient of the chemical potential force. Using the free electron gas chemical potential $\mu_0$, and assuming the deviation of $\mu$, $\delta \mu$ is very small, we have
 \begin{eqnarray}
 \mu=\mu_0 + \delta \mu, \quad \mu_0={\hbar^2 \over {2m_e}}(3 \pi^2 \rho_0)^{2/3}
 \label{eq-26mu}
 \end{eqnarray}
 Then, the value of $\delta \mu$ to the order of $\delta \rho$ is calculated as
  \begin{eqnarray}
 \delta \mu={ 2 \over 3} {\mu_0 \over \rho_0} \delta \rho
  \label{eq27}
 \end{eqnarray}
 
Including $\delta \mu$ and using Eq.~(\ref{eq27}),  Eq.~(\ref{eq24}) becomes 
\begin{eqnarray}
 { {\partial^2 \delta \rho}  \over {\partial t^2}} =-\Omega_{pl}^2\delta \rho+{{ 2 \mu_0} \over {3m_e}}\nabla^2 \delta \rho
 \end{eqnarray}
From this equation, the following dispersion relation is obtained for the Fourier component of $\delta \rho$ that oscillates with frequency $\Omega_{\bf k}$ and the wavenumber vector ${\bf k}$
\begin{eqnarray}
\Omega_{\bf k}^2=\Omega_{pl}^2 \left[ 1 + \left( {k \over k_{\rm TF}} \right)^2 \right]
 \label{eq29a}
 \end{eqnarray} 
 where the Thomas-Fermi wavenumber is defied by
 \begin{eqnarray}
k_{\rm TF}=\left( {{3e^2 \rho_0} \over {2 \epsilon_0 \mu_0}}
\right)^{1/2}
 \end{eqnarray} 
Although the constant factor for the second order term (it is $1$ in the present calculation) in Eq.~(\ref{eq29a}) is slightly different from the one obtained quantum mechanically using the random phase approximation (it is ${9 \over {10}}$; see
 \cite{Fetter-Walecka}), 
 essentially the same result is obtained.
 
Next, we consider the screening problem in the electron gas model.
 It is also a well-studied model; however, we consider it from the chemical potential view point.
  We follow the Pines' book \cite {Pines}. 
  Let us place a static charge $q$ at the origin of the electron gas.
 The electric field generated by it is expressed using the scalar potential $\varphi$, ${\bf E}=-\nabla \varphi$; it  satisfies the following equation
 \begin{eqnarray}
 -\epsilon_0 \nabla^2 \varphi=q \delta({\bf r})-e \delta \rho_{\rm ind}
 \label{eq31a}
 \end{eqnarray}
 where $\delta \rho_{\rm ind}$ denotes the induced charge density.
 We calculate $\delta \rho_{\rm ind}$ by employing the Thomas-Fermi model; then, the electron density is given by
 \begin{eqnarray}
 \rho({\bf r})={1 \over {3 \pi^2}} \left[ { {2m_e} \over \hbar^2} (\epsilon_{\rm F}+e \varphi  ) \right]^{3/2}
 \end{eqnarray}
 where $\epsilon_F$ is the Fermi energy.
 
From Eq.~(\ref{eq1mu}), the change of the chemical potential due to the addition of the point charge $q$ is given by
 \begin{eqnarray}
 \delta \mu_{\rm TF} =e\left[
\varphi + { \hbar \over {2e}}\partial_t (\delta \chi)
\right]
\label{eq2mu}
\end{eqnarray}
where the change of $\chi$ is also included.

Then, the electron density is given by
 \begin{eqnarray}
 \rho({\bf r})={1 \over {3 \pi^2}} \left[ {{2m_e} \over \hbar^2} (\epsilon_{\rm F}+\delta\epsilon_{\rm F} +\delta \mu_{\rm TF}) \right]^{3/2}, \quad \delta\epsilon_{\rm F}={ {\hbar} \over {2}}\partial_t (\delta \chi)
 \end{eqnarray}
Thus, $\delta \rho_{\rm ind}$ is approximately obtained as
  \begin{eqnarray}
 \delta \rho_{\rm ind}({\bf r})={{3  \rho_0} \over { 2  \mu_0}} \delta \mu_{\rm TF}
 \end{eqnarray}
  
 We rewrite the equation for $\varphi$ in Eq.~(\ref{eq31a}) to the equation for $\delta \mu_{\rm TF}$
   \begin{eqnarray}
 \nabla^2 \delta \mu_{\rm TF}=- {{eq} \over \epsilon_0} \delta({\bf r})+k_{\rm TF}^2\delta \mu_{\rm TF}
     \label{eq36}
 \end{eqnarray}
It yields
    \begin{eqnarray}
  \delta \mu_{\rm TF}= {{eq} \over { 4 \pi \epsilon_0 r}} e^{-k_{\rm TF} r} 
     \label{eq37}
 \end{eqnarray}
Thus, the scalar potential is obtained as
     \begin{eqnarray}
  \varphi= {{q} \over { 4 \pi \epsilon_0 r}} e^{-k_{\rm TF} r} -e^{-1}\delta \epsilon_F
     \label{eq38}
 \end{eqnarray}
Note that this contains the gauge ambiguity term $-e^{-1}\delta \epsilon_F$. 
 The value of $\epsilon_{\rm F}+\delta\epsilon_{\rm F}$ is obtained from the constraint that $\rho$ integrated over the system region is equal to the total number of electrons.
 
 The force from the electric field ${\bf E}$ is
 given by
\begin{eqnarray}
  -e{\bf E}= e \nabla \varphi= \nabla {{eq} \over { 4 \pi \epsilon_0 r}} e^{-k_{\rm TF} r}
\end{eqnarray}
 and that from the gradient of the chemical potential is
\begin{eqnarray}
  -\nabla \delta\mu_{\rm TF}=- \nabla {{eq} \over { 4 \pi \epsilon_0 r}} e^{-k_{\rm TF} r}
\end{eqnarray}
The sum of them is zero, indicating that the force acting on the electrons is zero.
Actually, it is a consequence of the fact that ${\bf E}$ is described only by the scalar potential.
In this case,  Eq.~(\ref{eq1mu}) yields the compensation of the electrostatic force and gradient of the chemical potential force if $\partial_t \chi$ dose not have the spatial dependence.

It is important to note that the force from the electrostatic field and that from the gradient of the chemical potential tend to compensate. This will occur in the situation where a steady electric current flows.

\section{Differentiation of normal and superconducting metals}
\label{Section3}

Before the advent of high transition temperature cuprate superconductors \cite{Muller1986}, it was believed that the BCS theory had solved all major problems of superconductivity \cite{BCS1957}. In this theory, the superconducting state is explained as
the state stabilized by electron-pair formation; the temperature where the electron pair is formed corresponds to the superconducting transition temperature.
Thus, the energy gap formed by the electron-pair formation was considered as the theoretical hallmark of superconductors, and methods to accurately obtain the energy gap formation temperature have been developed.

However, the elucidation of the mechanism of superconductivity in the cuprates has 
not been successful based on the BCS theory. A theory indicates that the superconducting transition temperature corresponds to the stabilization temperature of nano-sized loop currents \cite{Kivelson95}
and a simulation based on the spin-vortex-induced loop current model has yielded the result that supports this claim \cite{HKoizumi2015B}.
Besides, revisiting of experimental facts of superconductivity has found several problems even in the standard theory based on the BCS one:

\begin{enumerate}

\item The standard theory relies on  the use of particle number non-conserving formalism to explain the $U(1)$ gauge symmetry breaking although superconductivity occurs in an isolated system where the particle-number is conserved \cite{Peierls1991}.

\item The superconducting carrier mass obtained by the London moment experiment is the free electron mass $m_e$, although the standard theory prediction is the effective mass $m^\ast$ of the normal state \cite{Hirsch2013b}.
(The London moment is a phenomenon associated with the rotating superconductor.
A magnetic field is generated inside of it , and this magnetic field is called the `London magnetic field' and the magnetic moment associated with it is the `London moment', which has been measured.)

\item  The correct centripetal force in the London moment experiment cannot be obtained by the standard theory \cite{Hirsch2013b,HirschLorentz}. 

\item The reversible superconducting-normal phase transition in a magnetic field cannot be explained by the standard theory \cite{Hirsch2017}.

\item The dissipative quantum phase transition in a Josephson junction system predicted by the standard theory is absent \cite{PhysRevX2021a}.

\item 
The so-called `quasiparticle poisoning problem' indicates the existence a large amount of excited single electrons in Josephson junction systems for qubits, obtaining the observed ratio of their number to the Cooper pair number  $10^{-9} \sim 10^{-5}$ in disagreement with the standard theory ratio $10^{-52}$  \cite{poisoning2023,Serniak2019}.

\end{enumerate}

The existence of the above problems indicates the need for serious revisions of the superconductivity theory. A theory that encompasses the BCS theory has been developed by the present author using the Berry connection from many-electron wave functions \cite{koizumi2022,koizumi2023}
(A brief explanation of the relation between the new theory and the BCS one is given in \ref{Bogoliubov}). It resolves the above problems.
In the following, we explain how normal and superconducting metals are differentiated by the new theory. 

First, we consider a general case where the Berry connection from many-body wave functions is composed 
of contributions from a set of states $\{ \Psi_j \}$ whose occupation probability is given by the Boltzmann distribution,
\begin{eqnarray}
{\bf A}^{\rm MB}=\sum_j p_j {\bf A}_{\Psi_j}^{\rm MB}, \quad p_j={ {e^{- {E_j \over {k_B T}} }} \over  {\sum_j e^{-{ {E_j} \over {k_B T}} }}}
\label{multi-Berry-Conection}
\end{eqnarray}
where ${\bf A}_{\Psi_j}^{\rm MB}$ is the Berry connection from many-body wave functions for the wave function $\Psi_j$  \cite{Koizumi2024Lorentz,Koizumi_2023}, and $k_B$ is Planck's constant.
We rewrite ${\bf A}^{\rm MB}$ as
 \begin{eqnarray}
{\bf A}^{\rm MB}=-{1 \over 2} \nabla \chi
\label{eq-chi}
\end{eqnarray}
Here, we do not assume $\chi$ to be an angular variable with period $2\pi$ as in superconductors since it may be an average of ${\bf A}_{\Psi_j}^{\rm MB}$ from many $\Psi_j$'s. It becomes an angular variable with period $2\pi$ when only one $\Psi_j$ is involved. In any case, we express ${\bf v}$ in the form given in Eq.~(\ref{eq12v2}), and $\mu$ in Eq.~(\ref{eq1mu}). 

The supercurrent in Eq.~(\ref{eq2J}) explains a number of characteristics observed in superconductors as follows:

\begin{enumerate}

\item Persistent current: 

It can be explained as due to the presence of $\chi$ with stable non-zero $w[\chi]_C$ \cite{Koizumi_2023}.

\item Meissner effect: 

If $\nabla \times$ is operated on the both sides of Eq.~(\ref{eq2J}), $\nabla^2{\bf B}={{e^2 n} \over m_e}{\bf B}$ is obtained.
This is the equation obtained by London for explaining the Meissner effect.

\item Flux quantization: 

If we take a loop $C$ in Eq.~(\ref{eq2W}) along a path inside a ring-shaped superconductor where ${\bf B}={\bf j}=0$, we obtain $\oint_C {\bf A} \cdot d{\bf r}={h \over {2e}}w[\chi]_C$. This explains the observed
flux quantization.

\item London moment: 

Consider a rotating superconductor with an angular velocity ${\bm \omega}$. The velocity field inside the superconductor is ${\bf v}={\bm \omega}\times {\bf r}$. Substitute this ${\bf v}$ in Eq.~(\ref{eq12v2}), and operate $\nabla \times$ on the both sides of it, the relation $2{\bm \omega}={{e} \over m_e}{\bf B}$ is obtained. This indicates that the magnetic field ${\bf B}={{2m_e} \over e}{\bm \omega}$ is created inside the superconductor. This is the field that has been observed, experimentally.

\end{enumerate}

The normal conductor is the one without stable non-trivial $\chi$ that satisfies Eq.~(\ref{eq2Chi}).
In this case, the gauge ambiguity of ${\bf A}$ in Eq.~(\ref{eq2J}) is not absorbed by $\chi$. If ${\bf j}$ is calculated as an average value over different ${\bf A}$'s allowed by the gauge ambiguity, it will result in zero.

\section{Energy flow for a metallic wire connected to a battery}

Now we examine an old dissipation phenomenon, the Joule heating by electric current in a metallic wire connected to an electrical battery, with including the Berry connection from many-body wave functions.

The dissipation is believed to arise from the thermal fluctuation of the electric field.
We use the following Langevin equation \cite{VanKampen,QuantumNoise} to take into account the fluctuation
\begin{eqnarray}
m_e { {d{\bf v}} \over {dt}}=-{m_e \over \tau} {\bf v}+{\bm \eta}
\label{eqLangevin}
\end{eqnarray}
where $\tau$ is the relaxation time; the electron mass is simply taken to be the free electron mass, but the
effective mass $m^{\ast}$ will be more appropriate.  The thermal average of the noise ${\bm \eta}=(\eta_i, \eta_2, \eta_3)$ satisfies the following relations,
\begin{eqnarray}
\langle \eta_i(t) \rangle=0, \quad \langle \eta_i(t) \eta_j(t') \rangle={{2m_e k_B T} \over \tau} \delta_{i j } \delta(t-t')
\label{Langevin2}
\end{eqnarray}
where $k_B T$ arises from the assumption that the thermal average of velocity ${\bf v}=(v_1, v_2, v_3)$ satisfies
\begin{eqnarray}
{ 1 \over 2} m_e\langle v^2_i(t) \rangle={ 1 \over 2} k_B T
\label{equipartition}
\end{eqnarray}
according to the Maxwell velocity distribution.
The noise arising  from the fluctuating electric field
 is the cause of the dissipation. 

Now we consider the case where a force is added.
The Langevin equation in Eq.~(\ref{eqLangevin}) with an additional force ${\bf F}$ on the right-hand side has the following solution
\begin{eqnarray}
\langle {\bf v}(t) \rangle={ \tau \over m_e}{\bf F}(1-e^{-{t \over \tau}})+ \langle {\bf v}(0) \rangle e^{-{t \over \tau}}
\label{eq16}
\end{eqnarray}
which yields the final velocity
\begin{eqnarray}
\langle {\bf v}(\infty) \rangle={ \tau \over m_e}{\bf F}
\label{eqv-mu2}
\end{eqnarray}
The situation relevant to the Ohm's law is the one with the above stationary current, and we identify it to the drift velocity
\begin{eqnarray}
{\bf v}_d = { \tau \over m_e}{\bf F}
\label{drift-v}
\end{eqnarray}

When the stationary drift velocity is established, the velocity fluctuation occurs around ${\bf v}_d$. This situation may be described by modifying Eq.~(\ref{eqLangevin}) as
\begin{eqnarray}
m_e { {d({\bf v}-{\bf v}_d)} \over {dt}}=-{m_e \over \tau} ({\bf v}-{\bf v}_d)+{\bm \eta}', \quad  {\bm \eta}'={\bm \eta}+{\bf F}
\label{eqLangevin3}
\end{eqnarray}
with 
\begin{eqnarray}
\langle \eta_i'(t) \rangle_{\bf F}=0, \quad \langle \eta_i'(t) \eta_j'(t') \rangle_{\bf F}={{2m_e k_B T} \over \tau} \delta_{i j } \delta(t-t')
\label{eqLangevin4}
\end{eqnarray}
where $\langle \cdots \rangle_{\bf F}$ indicates the thermal average in the presence of ${\bf F}$.
Here, we assume that the effect of the presence of ${\bf F}$ on $\tau$ is negligible, 
and the same $\tau$ in Eq.~(\ref{eqLangevin}) can be used in Eq.~(\ref{eqLangevin3}).
The center of the noise is shifted, and the noise from it is denoted by ${\bm \eta}'$.

From Eqs.~(\ref{eqLangevin3}) and (\ref{eqLangevin4}), the following relation is obtained
\begin{eqnarray}
\langle {\bm \eta} \rangle_{\bf F}=-{\bf F}
\label{Langevin5}
\end{eqnarray}
It is the relation between the shift of the fluctuation force and applied one. 

Since the noise arises from the fluctuating electric field, the shift is the induced 
electric field given by
\begin{eqnarray}
{\bf E}_{\rm ind}={ 1 \over {-e}}\langle {\bm \eta} \rangle_{\bf F}
\label{eqLangevin6}
\end{eqnarray}
The steady current flows under the influence of the forces ${\bf F}$, ${\bf E}_{\rm ind}$, and the fluctuating
one  ${\bm \eta}'$.

\begin{figure}[H]
      \begin{center}
        \includegraphics[scale=0.35]{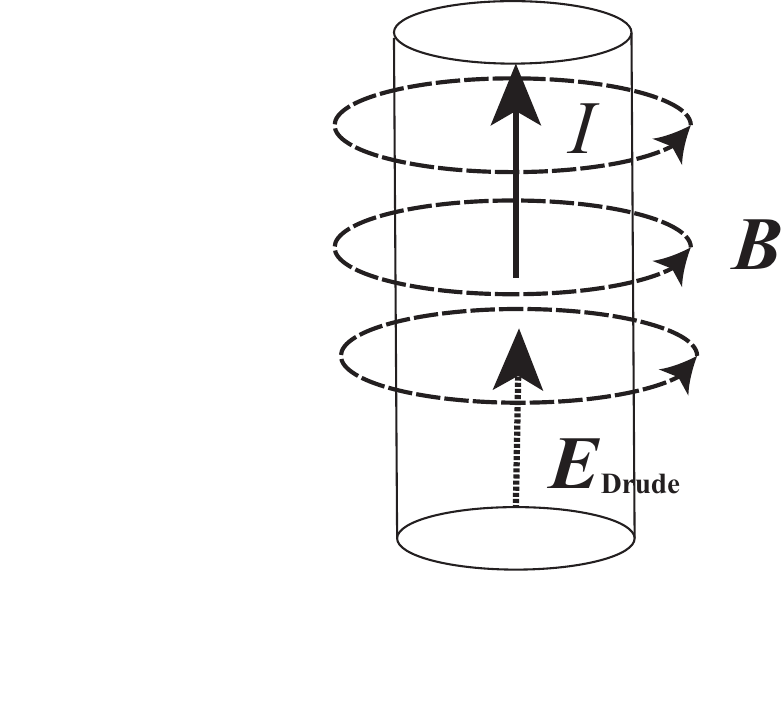}
      \end{center}
      \caption {A part of a metallic wire connected to a battery. The effect of the battery connection is the appearance of an electric field ${\bf E}_{\rm Drude}$ in the Drude model; this gives rise to the force that accelerates conduction electrons. The electric current $I$ flows in the same direction as ${\bf E}_{\rm Drude}$ in the wire, generating a circumferential magnetic field ${\bf B}$ around the wire. The Poynting vector ${\bf E}_{\rm Drude} \times {\bf B} /\mu_0$ points radially inwards to the wire. Thus,
      radiation enters into the wire. The radiation energy entered is consumed as the Joule heat ${\bf j} \cdot {\bf E}_{\rm Drude}$ in the wire.}
      \label{fig1A}
   \end{figure}
   
Now we consider the Joule heating by the Drude model.
In the textbook explanation based on the Drude theory,
the electric current is obtained using Eq.~(\ref{drift-v}) with ${\bf F}=-e{\bf E}_{\rm Drude}$.
The induced field in Eq.~(\ref{eqLangevin6}) is not taken into account, thus, the electric field is ${\bf E}_{\rm Drude}$. 
The voltage $V$ in Eq.~(\ref{Ohm}) is given by
\begin{eqnarray}
V=-\int_{{\bf r}_1}^{{\bf r}_2} {\bf E}_{\rm Drude} \cdot d {\bf r}
\end{eqnarray}
where ${{\bf r}_1}$ and ${{\bf r}_2} $ are points where the battery is connected.

The Poynting vector
${\bf E}_{\rm Drude}\times {\bf H}$ points radially inwards to the wire (see Fig.~\ref{fig1A}), supplying energy for the Joule heat ${\bf E}_{\rm Drude} \cdot {\bf j}=\sigma^{-1} {\bf j}^2$ \cite{FeynmanII27-5}.
 This is strange since it is sensible to consider the
energy is supplied by the battery by the current through the wire.

\begin{figure}[H]
      \begin{center}
        \includegraphics[scale=0.35]{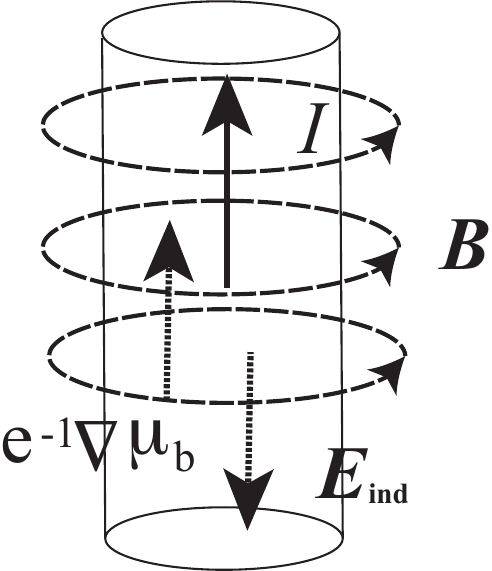}
      \end{center}
      \caption {The same as in Fig.~\ref{fig1A} but the battery connection gives rise to the chemical potential gradient $-\nabla \mu_b$ inside the wire. An electric field ${\bf E}_{\rm ind}$ is generated to counteract the
      force from the gradient of $-\nabla \mu_b$; due to the balance of the two forces, the electrons perform stationary translational motion with the velocity ${\bf v}_d$.
      The Poynting vector ${\bf E}_{\rm ind}\times {\bf B} /\mu_0$ points radially outwards from the wire, thus,
      radiation is emitted. The radiation energy emitted is equal to the Joule heat ${\bf j} \cdot {\bf E}_{\rm ind}$, which is the work done by the battery.}
      \label{fig1B}
   \end{figure}

Let us now assume that the voltage $V$ actually arises from the chemical potential difference realized by some chemical processes inside the battery. In other words, the electromotive fore is generated by chemical reactions on the electrodes, where electrons are pushed-in from  one of the electrodes, and pulled-out from the other.
 If the wire is made of a simple metal, the Fermi level increases on the number density increased connection point, and decreases on the other (see Eq.~(\ref{eq-26mu})). This difference gives rise to the electromotive force that generates the current.
Then, the voltage $V$ is given by
\begin{eqnarray}
V=-e^{-1} \int_{{\bf r}_1}^{{\bf r}_2} \nabla \mu_b \cdot d {\bf r}=-{{\mu_b({\bf r}_2) - \mu_b({\bf r}_1)} \over e}
\label{eqV-1}
\end{eqnarray}
where $\mu_b$ is the chemical potential whose origin is the battery.

We consider that the situation given by the Langevin equation in Eq.~(\ref{eqLangevin3}) is relevant,
rather than the one described by the Langevin equation in Eq.~(\ref{eqLangevin}).
In this case, as in the screening problem considered in Sec.\ref{Plasma}, the balance of
the electric force and the gradient of the chemical potential force is taken into account.
By taking ${\bf F}=-\nabla \mu_b$, and using Eqs.~(\ref{drift-v}), (\ref{Langevin5}), and (\ref{eqLangevin6}), 
the following relations are obtained, 
\begin{eqnarray}
{\bf j}= e^{-1} \sigma \nabla \mu_b, \quad 
{\bf E}_{\rm ind}=-e^{-1}\nabla \mu_b
\label{Langevin7}
\end{eqnarray}
where $\sigma$ is given by
\begin{eqnarray}
\sigma={{e^2 n_e \tau} \over m_e}
\end{eqnarray}
with $n_e$ being the conduction electron density.
The direction of ${\bf E}_{\rm ind}$ is opposite to the current direction (see Fig.~\ref{fig1B}); thus, the direction of the Poynting vector ${\bf E}_{\rm ind}\times {\bf B} /\mu_0$ points radially outwards from the wire. 

If we follow the typical textbook calculation of the Poynting theorem energy flow for the above situation, it goes as follows:
Let us assume the part of the wire depicted is a cylindrical shape with circular cross sectional area of radius $r$
and length $\ell$. Then, a magnetic field with the magnitude $H ={I \over {2 \pi r}}$ exists at the surface of the wire. The magnitude of the Poynting vector ${\bf E}_{\rm ind}\times {\bf H} $ is ${{IE_{\rm ind}} \over {2 \pi r}}$.
By integrating it over the side surface of the cylinder, the total energy flow is obtained as
\begin{eqnarray}
\int_{\cal S} d{\bf S} \cdot ({\bf E}_{\rm ind} \times {\bf H})=IE_{\rm ind}\ell
\label{eq28}
\end{eqnarray}
where ${\cal S}$ is the surface of the cylinder; the Poynting vector goes out only through the side surface of the cylinder 
with area $2\pi r$.
Since ${\bf E}_{\rm ind}$ points in the opposite direction to ${\bf j}$, we have
\begin{eqnarray}
\int_{\cal V} d^3r \  {\bf j}  \cdot  {\bf E}_{\rm ind}=-\int_{\cal V} d^3r \  j E_{\rm ind}=-{IE_{\rm ind} \ell}
\label{eq29}
\end{eqnarray}
where ${\cal V}$ is the volume of the cylinder, and relation  $I=j \pi r^2$ is used. Then, the relation
in Eq.~(\ref{eqPoynting}) is satisfied since the electromagnetic field energy in the wire is constant
in the stationary situation.
The term with ${\bf j} \cdot {\bf E}_{\rm ind}$ has the minus sign, indicating that it is the work done by the battery. 
Thus, the Poynting theorem now states that 
the energy for the emitted radiation is supplied by the battery through the wire. This is a sensible result. This emitted radiation is the Joule heat; when it is absorbed by some materials it will cause heating.

Actually, the Drude model description for the energy flow is relevant for the optical conductivity case. In this case the electric field is applied from outside of the sample as radiation, and
the relation in Eq.~(\ref{eqOhm}) is valid. The absorption of the radiation occurs in the sample, which can be calculated usually using the Fermi's golden rule formalism of quantum mechanics;
this will correspond to the radiation energy consumption. 

 The conductivity by the battery connection is, in a sense, the reverse process of the optical conductivity case. For the battery connection case, collisions among the conduction electrons and those between the ion cores and the electrons generate quantum mechanical excited states that correspond to the exited states created by the absorption of the radiation in the optical conductivity case.
The emission of radiation is due to the relaxation of the excited states, which is the reverse process of the excitation by the applied radiation for the optical conductivity case.
 The generation of the induced field ${\bf E}_{\rm ind}$ and resulting emission of radiations by the Poynting vector may be considered as classical approximations for the above quantum mechanical processes. 
 
The differences between metals and semiconductors cannot be described by the present method. It requires microscopic quantum mechanical calculations, 
which can differentiate electronic excitations in  metals and semiconductors. 

\section{Energy flow during discharging of a capacitor}

\begin{figure}[H]
      \begin{center}
        \includegraphics[scale=0.4]{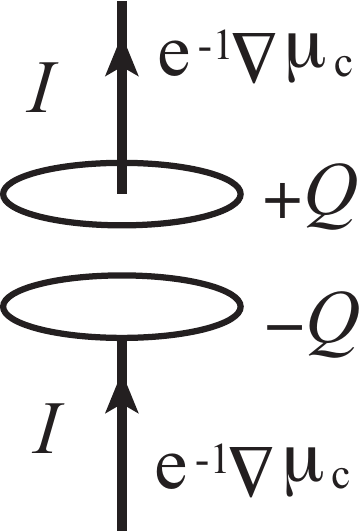}
      \end{center}
      \caption {A capacitor of capacitance $C$ with charge $Q$ stored and a wire with resistance $R$ that shunts the capacitor. During discharging $Q$ decreases as $Q(t)=Q_0 e^{-{1 \over {RC}}t}$,
      where $Q_0$ is the initial charge at $t=0$. The electric current $I={{dQ} \over {dt}}$ flows after $t=0$.
      We assume the balance of the voltage given in Eq.~(\ref{eqbalance2}) is maintained in the quasi-stationary current situation.
      The electric field energy originally stored between the electrodes is ${{Q_0^2} \over {2C}}$,
      which is emitted as the radiation from the wire.}
      \label{fig2}
   \end{figure}

Let us consider the discharging of a capacitor. In the beginning, the charges $+Q_0$ and $-Q_0$ are on each 
electrodes of the capacitor, respectively (see Fig.~\ref{fig2}), and the chemical potential difference $Q_0/C$ exists across the electrodes, where $C$ is the capacitance of the capacitor. 
This chemical potential difference is due to the difference of the electron number density on each electrode. 
We may consider that this situation is realized first by connecting the capacitor to a battery of the voltage $Q_0/C$, and then, by disconnecting the battery after the charging that establishes the voltage balance between the battery and the capacitor.

The energy stored as the energy of the electric field ${\bf E}$ between the electrodes.
The balance of the voltage between the chemical potential difference between the two electrodes and
the electromagnetic field ${\bf E}$ in the space between the electrodes is given by
\begin{eqnarray}
 V=-\int_1^2 {\bf E} \cdot d{\bm \ell}=
 {1 \over {-e}} \int_1^2 \nabla \mu_c \cdot d{\bm \ell}
 \label{eqbalance2}
 \end{eqnarray}
where `$1$' and `$2$' indicate points in the electrodes; the subscript $c$ in the chemical potential $\mu_c$ indicates that it is the capacitor origin. 

Now, we shunt the capacitor by the connection of a resistive wire. We assume a quasi-stationary current flow. It flows due to the chemical potential difference between the electrodes. The capacitor acts as the battery that produces the gradient of
the chemical potential $\nabla \mu_c$.
The decrease of the electric field ${\bf E}$ occurs due to the decrease of the charge stored on the capacitor, thus, the reduction of the electric field energy stored also occurs. A typical textbook explanation goes  as follows: The time variation of ${\bf E}$ generates a magnetic field ${\bf B}$ by the displacement current,
and the Poynting vector ${\bf E} \times {\bf H}$ directing radially outwards from the capacitor is generated. 
The reduction of the electric field energy occurs due to the outward radiation from the space between the electrodes. 

On the other hand, the present explanation is as follows: ${\bf A}$ and $\varphi$ are fundamental physical quantities instead of ${\bf E}$ and ${\bf B}$; then, ${\bf E}$ and ${\bf B}$ are two disguises of ${\bf A}$ and ${\varphi}$.
If we adopt the gauge $\nabla \cdot {\bf A}=0$, the vector potential is given as follows \cite{FeynmanII21-3}
\begin{eqnarray}
{\bf A}({\bf r}, t)={\mu_0 \over {4\pi}}\int d^3 r'{ {{\bf j}({\bf r}', t-|{\bf r}-{\bf r}'|/c)}
\over {|{\bf r}-{\bf r}'|}}
\end{eqnarray}
This indicates that only the true current ${\bf j}$ generates ${\bf A}$; thus, the generation of ${\bf B}$
by the displacement current does not occur. It may be worth pointing out that
Maxwell obtained a similar formula (the vector potential is called the `electromagnetic momentum') \cite{Maxwell-617}; he included the displacement current in ${\bf j}$, which is incorrect.

The energy flow occurs through the wire by the current generated by the chemical potential gradient generated by the capacitor that acts as a battery.
Applying Eq.~(\ref{eq29}) for the wire in Fig.~\ref{fig2} with neglecting the effect of bending of the wire,
the rate of the work done by the capacitor is calculated as
\begin{eqnarray}
\int_{\cal V} d^3r \  {\bf j}  \cdot  {\bf E}={I(t)V(t)}
\label{eq31}
\end{eqnarray}
where 
$V(t)$ is the voltage on the capacitor, $I(t)$ is the current flowing in the wire, and now ${\cal V}$ is the volume of the shunt wire.
The total work done by the capacitor is obtained by integrating Eq.~(\ref{eq31}) in the time interval $0 \leq t \leq \infty$ as
\begin{eqnarray}
\int_0^\infty I(t)V(t) dt=\int_0^\infty {{dQ(t)} \over {dt}} {{Q(t)} \over C }dt=-{ Q_0^2 \over {2C}}
\end{eqnarray}
where $Q(t)=Q_0 e^{-{1 \over {RC}}t}$ is used. The total work capacitor done is equal to the energy stored originally as the electric field. As shown in Eq.~(\ref{eq28}), this energy is equal to the energy irradiated from the wire as the Joule heat.

\section{Josephson junction}

\begin{figure}[H]
      \begin{center}
        \includegraphics[scale=0.8]{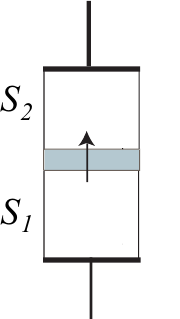}
      \end{center}
      \caption {Josephson junction composed of two superconducting electrodes ($S_1$ and $S_2$) and an insulator (gray region) between them.
      This structure inherently has a capacitor contribution due to the presence of two electrodes facing each other with the insulator between them. }
      \label{figJJ}
   \end{figure}
   
Now we consider the supercurrent flow through the Josephson junction.
The Josephson junction is schematically depicted in Fig.~\ref{figJJ}.
The current through it is given by
   \begin{eqnarray}
I=I_c \sin \phi
\label{eq-varphi}
\end{eqnarray}
where $I_c$ is the critical current of the junction and $\phi$ is the superconducting phase difference between the 
two superconductors \cite{FeynmanIII21-9}. When a finite voltage exits across the Josephson junction, the time variation of $\phi$ given by
\begin{eqnarray}
{ {d \phi} \over {dt}}={{2eV} \over \hbar}
\label{J-R}
\end{eqnarray}
occurs,
where $V$ is the voltage across the junction.  This formula has been confirmed, experimentally \cite{Shapiro63}. It was first obtained by Josephson \cite{Josephson62}, and also derived by Feynman in his textbook \cite{FeynmanIII21-9}.
By assuming the Cooper-pair tunneling, $\phi$ is given by
\begin{eqnarray}
\phi={{2e} \over \hbar} \int_1^2\left({\bf A}-{{\hbar} \over {2e}}\nabla  \chi \right) \cdot d{\bf r}
\label{eqphi1}
\end{eqnarray}
where `$1$' and `$2$' are points in the two superconductors $S_1$ and $S_2$ of the junction, respectively.
Note that $\left({\bf A}-{{\hbar} \over {2e}}\nabla  \chi \right)$ is gauge invariant, and $2e$ in the factor ${{2e} \over \hbar}$ is the Cooper-pair charge.

\begin{figure}[H]
  \begin{center}
        \includegraphics[scale=1.]{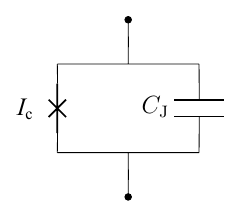}
      \end{center}
	\caption{A Josephson junction composed of a capacitor with capacitance $C_{\rm J}$ and an electron transfer part with critical current $I_{\rm c}$.}
	\label{JJ2}
\end{figure}

A Josephson junction is composed of an electron transfer contribution and a capacitor contribution as depicted with circuit elements in Fig.~\ref{JJ2}. The capacitor contribution arises from the structure of two facing electrodes as seen in Fig.~\ref{figJJ}.
The tunneling electron goes through the transfer part with experiencing the electric field between the superconducting electrodes giving rise to the capacitor contribution. This capacitor contribution is lacking in the Josephson's and textbook derivations \cite{Josephson62,FeynmanIII21-9}.

Let us calculate the time derivative of Eq.~(\ref{eqphi1}) with including the capacitor contribution. The result is
\begin{eqnarray}
{{ d\phi} \over {dt}}&=&{{2e} \over \hbar} \int_1^2\left(\partial_t{\bf A}-{{\hbar} \over {2e}}\nabla  \partial_t\chi \right) \cdot d{\bf r}
\nonumber
\\
&=&{{2e} \over \hbar} \int_1^2\left(-{\bf E}-e^{-1}\nabla \mu \right) \cdot d{\bf r}
\label{eqphi2}
\end{eqnarray}
where $\mu$ in Eq.~(\ref{eq1mu}) and  ${\bf E}=-\nabla \varphi-\partial_t {\bf A}$ are used.
The voltage balance between the gradient of the chemical potential part and
the electric field part 
gives rise to the relation in Eq.~(\ref{eqbalance2}).
Then, we have ${ {d \phi} \over {dt}}={{4eV} \over \hbar}$, which disagrees with Eq.~(\ref{J-R}).
This discrepancy is resolved if we consider that the electron transfer occurs singly, not pair-wise. This indicates that 
the factor ${{2e} \over \hbar}$ in Eq.~(\ref{eqphi1}) is actually ${{e} \over \hbar}$. This situation was considered by the present author, previously \cite{koizumi2021b}.
It is also worth noting that Eq.~(\ref{single-valued3}) indicates that the
single-electron change in the superconducting condensate in the new theory.

Since the pairing stabilization exists in superconductors, there are situations where subsequent single electron transfers observed as
the pair-wise transfers. Thus, this result does not contradict experiment.
It is noteworthy that it explains the so-called `quasiparticle poisoning problem': In quantum electronic circuits with Josephson junctions, a large amount of excited single electrons in Josephson junctions is observed, with the observed ratio of their number to the Cooper pair number  $10^{-9} \sim 10^{-5}$ in disagreement with the standard theory ratio $10^{-52}$\cite{poisoning2023,Serniak2019}; this may be the consequence of abundant single-electron transfer.

\section{Planckian dissipation}

We consider another superconductivity related problem. It is so-called, the `Planckian dissipation' problem.
The dissipations with the relaxation time of the Planckian time order have been observed in anomalous metals \cite{Sachev1992,Zaanen2019,Planckian}, including the high transition temperature cuprate superconductors. It is believed that the elucidation of it is the key to understand the cuprate superconductivity.
We examine this problem with including the fluctuation of the gradient of the chemical potential force.

First, we consider the fluctuation of $-e \nabla\varphi$ term arising from $-\nabla \mu$, where $\mu$ is given in Eq.~(\ref{eq1mu}).
We assume  that the fluctuation from  $-e\nabla \varphi$ can be mimicked by the force from an  electric field $e{\bf E}$.
The motion of electrons is assumed in the $x$ direction.
Then, we have
\begin{eqnarray}
\tau ={{2m_e k_B T} \over {e^2\langle E_x(t) E_x(t') \rangle}} \delta(t-t')\approx {{2m_e k_B T} \over {e^2\langle E_x(t) E_x(t) \rangle}} {1 \over \tau_E}
\label{Langevin-Tau}
\end{eqnarray}
where $E_x$ is the $x$ component of the fluctuating electric field. The effect of the delta function is replaced by a temperature independent constant $\tau_E$ with the dimension of time. This is the crudest approximation based on the following:
$\langle E_x(t) E_x(t') \rangle$ is assumed to be non-zero in the time interval $\tau_E$ with a
typical value $\langle E_x(t) E_x(t) \rangle$, and the delta function is replaced by a rectangle with the value $\tau_E^{-1}$ in the time-interval $\tau_E$.

We may express the electric field  $E_x(t)$ using the creation and annihilation operators for the photon with energy $\hbar \omega_{n}$, $\hat{a}_{k_n}^\dagger$ and $\hat{a}_{k_n}$, respectively
\begin{eqnarray}
E_x (t)=i\sum_n \sqrt{{\hbar \omega_n} \over {2 \epsilon_0 L_x L_y L_z}}\left( \hat{a}_{k_n} e^{-i(\omega_n t- k_n x)}- \hat{a}_{k_n}^\dagger e^{i(\omega_n t- k_n x)}
\right)
\end{eqnarray}
 $L_xL_yL_z$ is the volume of a rectangular cuboid with length $L_x$ in the $x$ direction, width $L_y$ in the $y$ direction, and height $L_z$ in the $z$ direction; the wave number $k_n$ is given by $k_n={{2 \pi} \over L_x}n, \
 \mbox{($n$ is an integer)}$, and $\omega_n$ by $\omega_n = c |k_n|$.

Then, $\langle E_x(t) E_x(t) \rangle$ is estimated as
 \begin{eqnarray}
\langle E_x(t) E_x(t) \rangle &=& \sum_n{  {\hbar \omega_n} \over {2 \epsilon_0 L_x L_y L_z}} \langle 2 \hat{a}^\dagger_{k_n} \hat{a}_{k_n}+1 \rangle
\nonumber
\\
&\approx &{  {1} \over {\pi \epsilon_0 L_y L_z}}
 \int_0^{\infty} dk \ \hbar c k e^{-{{\hbar c k}\over {k_B T}}}
 \nonumber
 \\
 &=&{ { \hbar c} \over {\pi \epsilon_0 L_y L_z}}
\left(  {{k_B T} \over {\hbar c}} \right)^2
\end{eqnarray}
where the sum $\sum_n$ is replaced by ${L_x \over {2\pi} } \int dk$ and the approximation $\langle 2 \hat{a}^\dagger_{k_n} \hat{a}_{k_n}+1 \rangle \approx 2 e^{-{{\hbar c |k_n|}\over {k_B T}}}$ is used, where $e^{-{{\hbar c |k_n|}\over {k_B T}}}$ is the Boltzmann factor.
Then, $\tau$ is given by
\begin{eqnarray}
\tau \approx {{m_e L_y L_z } \over {2\alpha \hbar \tau_E }} {\hbar \over {k_B T}}
\label{Langevin-Tau1}
\end{eqnarray}
where $\alpha \approx {1 \over 137}$ is the fine structure constant.
There are unknown parameters, $L_y$, $L_z$ and $\tau_\chi$, but 
$\tau$ exhibits the dependence on ${\hbar \over {k_B T}}$, a signature of the Planckian dissipation.
Due to the presence of $\alpha^{-1}$, however, the above estimate is too large to explain the observed $\tau \sim  {\hbar \over {k_B T}}$ relation if unknown parameters are in the order of $1$ in the atomic units.

Let us now estimate $\tau$ using the fluctuation of $-{\hbar \over 2}\partial_t(\nabla \chi)$ from $-\nabla \mu$, the `gauge fluctuation'; we assume 
it is 
associated with the fluctuation of the winding number in Eq.~(\ref{eq2W}). First, we write $\tau$ as
\begin{eqnarray}
\tau = {{2m_e k_B T} \over {{\hbar^2 \over 4}\langle \partial_t \partial_x \chi(t) \partial_t \partial_x \chi(t')  \rangle}} \delta(t-t')
\approx {{2m_e k_B T} \over { {\hbar^2 \over 4}\langle \partial_t \partial_x \chi(t) \partial_t \partial_x \chi(t)  \rangle}} {1 \over \tau_\chi}
\label{Langevin-Tau2}
\end{eqnarray}
where the effect of the delta function is replaced by a temperature independent constant $\tau_E$ with the dimension of time as in the previous case.

We also perform the following approximation
\begin{eqnarray}
\partial_t \partial_x \chi(t) \approx {{\partial_x \chi(t+\tau_\chi) - \partial_x \chi(t)} \over \tau_\chi}
\approx {{\partial_x \chi(t+\tau_\chi)} \over \tau_\chi}, \mbox{  or  } {{ - \partial_x \chi(t)} \over \tau_\chi}
\end{eqnarray}
This approximation is based on the assumption that the abrupt change of $\partial_x \chi$ occurs in an average interval of $\tau_\chi$, and the abrupt change is due to the creation or annihilation of the vortices of $\nabla \chi$
that is approximated as $\partial_x \chi =0$ in the before or in the after. We assume such abrupt change occurs as the consequence of the fluctuation of the winding number in Eq.~(\ref{eq2W}).

Using the above approximation and assuming that the vortices exit in the $xy$ plane, we have
\begin{eqnarray}
 \langle \partial_t \partial_x \chi(t) \partial_t \partial_x \chi(t)  \rangle
\approx { 1 \over \tau^2_\chi} \langle \partial_x \chi(t)  \partial_x \chi(t)  \rangle
\approx  { 1 \over {2 \tau^2_\chi} }\langle (\nabla \chi(t))^2  \rangle
\label{Langevin-Tau3}
\end{eqnarray}

Next, we use the following approximation for the thermal average,
\begin{eqnarray}
{\hbar^2 \over {8 m_e}} \langle (\nabla \chi(t))^2 \rangle \approx 2{ 1 \over {\Delta E_\chi}} \int_0^\infty dE_\chi E_\chi e^{-{E_\chi \over {k_B T}}}
={ 2 \over {\Delta E_\chi}} (k_B T)^2
\label{Langevin-Tau4}
\end{eqnarray}
where $E_\chi={\hbar^2 \over {8 m_e}} (\nabla \chi(t))^2$ is used for the energy for the vortex of $\chi$; it is the kinetic energy of mass $m_e$ and the velocity ${\hbar \over {2m_e}} \nabla \chi$ given in Eq.~(\ref{eq12v2}) with neglecting ${\bf A}$. ${\Delta E_\chi}$ is the typical energy fluctuation caused by the gauge fluctuation; and factor $2$ takes into account the degeneracy due to the state with $\nabla \chi$ and that with $-\nabla \chi$.

Overall, $\tau$ is given by
\begin{eqnarray}
\tau 
&\approx& {{ \tau_\chi \Delta E_\chi} \over { k_B T}}
\approx {{ \hbar} \over { k_B T}}
\label{Langevin-Tau5}
\end{eqnarray}
where $\tau_\chi \Delta E_\chi \approx \hbar$ is used in the last equality. This explains the Planckian time-scale.

The above Planckian time behavior was observed in the cuprate superconductor 
Bi$_2$Sr$_2$CuO$_{6+\delta}$ from the superconducting transition temperature T$_c$ $\approx$ 7 up to 700 K \cite{PhysRevB.41.846}. 
The superconductivity in the cuprate occurs in the CuO$_2$ plane \cite{Jiang:2014aa,Yu:2019aa}; and an estimate of the transition temperature indicates it corresponds to the stabilization of nanometer-sized loop currents in that plane \cite{Kivelson95}. The BKT transition type superconducting transition temperature behavior in the cuprate also indicates it is related to the stabilization of loop currents \cite{D3RA02701E}.
 A computer simulation for a model for loop currents in the cuprate exhibits that the superconducting transition 
 temperature corresponds to the temperature where the fluctuation of $-{\hbar \over 2}\partial_t(\nabla \chi)$ 
 is suppressed \cite{HKoizumi2015B}. Taking into account all the results, it is highly plausible to consider that the Planckian time behavior of the relaxation time 
 in the cuprate superconductivity is related to the superconducting mechanism of it, where the supercurrent is generated as a collection of stable loop currents of the nanometer size in the CuO$_2$ plane.

\section{Concluding remarks}

In textbooks, the Ohm's law is explained by assuming that the effect of the battery connection is to generate electric field inside the metallic wire. During the development of the Drude theory, the force from the gradient of chemical potential was
not known; thus, the forces used were the electric field force and the friction force. By considering the
balance of them, the conductivity was obtained. 
Since the friction force causes the energy dissipation, the Joule heat was attributed to it.
However, this theory is known to give an odd energy flow when the Poynting theorem is used \cite{FeynmanII27-5,Poynting2022}. It is also notable that the
steady current state appears after the drift velocity becomes constant; thus, how this velocity is maintained needs to be considered; in the present theory, it is maintained by the shift of the fluctuation center and the appearance of the induced electric field given in Eqs.~(\ref{Langevin5}) and (\ref{eqLangevin6}). On the other hand, the Drude theory only considers how this drift velocity is achieved as shown in  Eqs.~(\ref{eqv-mu2}) and (\ref{drift-v}).
Note that the Drude theory is applicable for the optical conductivity case, where the radiation enters from outside, and absorbed by the wire.

The chemical potential employed in this work is not the thermodynamical origin, in contrast to the one introduced by
Gibbs \cite{Gibbs}. In the electron gas theory by Sommerfeld \cite{Sommerfeld:1928aa}, the chemical potential appears in
the Fermi distribution function. In the more elaborated Boltzmann equation method, the spatial variation of the chemical potential is taken into account, and the chemical potential gradient force appears. 
However it starts from the free electron model; on the other hand, the present method starts from the many-body wave function,
thus, the many-body effect is included from the beginning. 

For the explanation of the Planckian dissipation problem, `gauge fluctuation', 
the fluctuation of $-{\hbar \over 2}\partial_t(\nabla \chi)$, is the key ingredient.
If this fluctuation is suppressed $({\bf A}-{\hbar \over {2e}}\nabla \chi)$ becomes gauge invariant; then, ${\bf v}$ that explicitly depends on ${\bf A}$ is realized.
Conversely, we may say that the reason for the gauge invariance in ${\bf E}$ and ${\bf B}$ in the classical electromagnetic theory is due to the fact that it deals with cases where the `gauge fluctuation' is so large that physical
variables are those survive the noise from it; and ${\bf E}$ and ${\bf B}$ are such variables.

In Maxwell's equations, ${\bf A}$ and $\varphi$ do not appear, although Maxwell used them to formulate the equations \cite{Maxwell-618}. Most notably, he obtained the electromagnetic wave using the vector potential \cite{Maxwell-784}.
Feynman stated, `If we take away the scaffolding he used to build it, we find that Maxwell's beautiful edifice stands on its own' \cite{FeynmanII18-1}. 
The present work indicates the possible revision of this statement may be needed: What used to be considered as the scaffolding is actually part of the edifice; the noise from `gauge fluctuation' blurs its appearance, and we have been mistakenly consider the part made of ${\bf E}$ and ${\bf B}$ exhibits the whole beauty of the edifice. If the noise is cleared, the true beauty of the edifice is revealed. It is made of ${\bf A}$ and $\varphi$ with accompanying the Berry connection. 

\vspace{10mm}

{\bf Author contributions}: H.K. contributed all.

\vspace{10mm}

{\bf Competing interests. }: The author declare that there are no competing interests in this work.

\vspace{10mm}

\begin{appendices}

\section{The relation between the BCS theory and the new theory using the Berry connection}
\label{Bogoliubov}

For the convenience of the reader, we succinctly explain the relation between the BCS theory and the new theory that encompasses it. More detailed explanation may be found in Refs.~\cite{koizumi2023,koizumi2021,Taya2025}.

The BCS superconducting state vector is given by
\begin{eqnarray}
|{\rm BCS} (\theta) \rangle=\prod_{\bf k}\left(u_{\bf k}+v_{\bf k}c^{\dagger}_{{\bf k} \uparrow}c^{\dagger}_{-{\bf k} \downarrow}
e^{ {i}{\theta}} \right)|{\rm vac} \rangle
\label{BCS}
\end{eqnarray}
where real parameters $u_{\bf k}$ and $v_{\bf k}$ satisfy $u_{\bf k}^2+v_{\bf k}^2=1$.
It is a linear combination of different particle number states, thus, breaks the global $U(1)$ gauge invariance. The spatial variation of $\theta$ (which is absent in Eq.~(\ref{BCS}) but appears later in 
Eq.~(\ref{field2p})) is the Nambu-Goldstone mode and
the vector potential ${\bf A}$ appears in the theory as the following gauge invariant combination
\begin{eqnarray}
{\bf A}+{\hbar \over {2e}}\nabla \theta
\label{theta-Nambu}
\end{eqnarray}

The BCS theory is systematically described using the following 
Bogoliubov operators \cite{Bogoliubov58} given by
\begin{eqnarray}
 \gamma^{\rm BCS}_{{\bf k} \uparrow }&=& u_{k} e^{-{i \over 2}\theta} c_{{\bf k} \uparrow} -  v_{ k} e^{{i \over 2}\theta} c^{\dagger}_{-{\bf k} \downarrow}
\nonumber
\\
\gamma^{\rm BCS}_{-{\bf k} \downarrow }&=& u_{k} e^{-{i \over 2}\theta} c_{-{\bf k} \downarrow} +  v_{ k} e^{{i \over 2}\theta} c^{\dagger}_{{\bf k} \uparrow}
\label{gamma-BCS}
\end{eqnarray}
The superconducting ground state is defined as the ground state for the Bogoliubov excitations
\begin{eqnarray}
 \gamma^{\rm BCS}_{{\bf k} \uparrow }|{\rm BCS} \rangle=0, \quad \gamma^{\rm BCS}_{-{\bf k} \downarrow }|{\rm BCS} \rangle=0
 \label{gamma-BCS2}
 \end{eqnarray}
 
The difference of the field operators in the normal and superconducting states is crucial.
The normal metallic state is assumed to be well-described by the free electrons with the effective mass $m^{\ast}$.
 Then, the electron field operators are given by
\begin{eqnarray}
\hat{\Psi}^{\rm norm}_{\sigma}({\bf r})={ 1 \over \sqrt{V}}\sum_{\bf k} e^{i {\bf k} \cdot {\bf r}} c_{{\bf k} \sigma}
\end{eqnarray}
On the other hand, in the superconducting state, the electron field operators are expressed 
using the Bogoliubov operators as follows
\begin{eqnarray}
\hat{\Psi}^{\rm BCS}_{\uparrow}({\bf r})&=& \sum_{\bf k} e^{{i \over 2}\theta} \left[ \gamma^{\rm BCS}_{{\bf k} \uparrow } u_{\bf k}({\bf r}) -  (\gamma^{\rm BCS})^{\dagger}_{-{\bf k} \downarrow } v_{\bf k}({\bf r}) \right]
\nonumber
\\
\hat{\Psi}^{\rm BCS}_{\downarrow}({\bf r})&=& \sum_{\bf k} e^{{i \over 2}\theta}\left[ \gamma^{\rm BCS}_{{\bf k} \downarrow } u_{\bf k}({\bf r}) + (\gamma^{\rm BCS})^{\dagger}_{-{\bf k} \uparrow } v_{\bf k}({\bf r}) \right]
\label{fieldOp}
\end{eqnarray}
where the following coordinate dependent functions are introduced
\begin{eqnarray}
u_{\bf k}({\bf r})={1 \over \sqrt{V}}e^{ i {\bf k} \cdot {\bf r}}u_k, \quad
v_{\bf k}({\bf r})={1 \over \sqrt{V}}e^{i {\bf k} \cdot {\bf r}}v_k
\end{eqnarray}

For considering the case where the coordinate dependent functions are different from plane waves,
we use the label $n$ in place of the wave number ${\bf k}$. Then, the field operators become
 \begin{eqnarray}
\hat{\Psi}^{\rm BCS}_{\uparrow}({\bf r})&=&\sum_{n} e^{{i \over 2}\theta({\bf r}) }\left[ \gamma^{\rm BCS}_{{n} \uparrow } u_{n}({\bf r})  -(\gamma^{\rm BCS})^{\dagger}_{{n} \downarrow } v^{\ast}_{n}({\bf r}) \right]
\nonumber
\\
\hat{\Psi}^{\rm BCS}_{\downarrow}({\bf r})&=&
\sum_{n} e^{{i \over 2}\theta({\bf r})} \left[ \gamma^{\rm BCS}_{{n} \downarrow } u_{n}({\bf r}) + (\gamma^{\rm BCS})^{\dagger}_{{n} \uparrow } v^{\ast}_{n}({\bf r}) \right]
\label{field2p}
\end{eqnarray}
where the coordinate dependent $\theta$, $\theta({\bf r})$, is also introduced.
With this $\theta({\bf r})$, the combination in Eq.~(\ref{theta-Nambu}) becomes gauge invariant.
The formalism that uses the above field operators is the essence of the Bogoliubov-de Gennes formalism \cite{deGennes,Zhu2016}.
At this point, we may view the superconducting state in the standard theory
is the theory that goes beyond the Schr\"{o}dinger's wave mechanics by breaking the global $U(1)$ gauge invariance; and the phase factor $e^{{i \over 2}\theta({\bf r})}$ plays the important role to produce the supercurrent \cite{Anderson,Weinberg}.
 
The new theory is constructed by comparing Eqs.~(\ref{single-valued3}) and (\ref{field2p}); we notice that $e^{{i \over 2}\theta({\bf r})}$ may describe the phase factor with the Berry connection.
Then, the following correspondence may be possible
\begin{eqnarray}
\theta \rightarrow -\hat{\chi}
\label{eqA9}
\end{eqnarray}
where $\hat{\chi}$ is the operator version of $\chi$. Here, we use the operator version of $\chi$, $\hat{\chi}$, is used so that the change of the number of electrons participate in the collective mode described by $\chi$ (we call it, the `$\chi$ mode') is possible.

The $\chi$ mode produces supercurrent in a similar manner as $\theta({\bf r})$ does. When it is quantized its canonical conjugate variable is the number of particles participating in it as shown below: In order to find the conjugate variable of $\chi$,
we use the following Lagrangian 
\begin{eqnarray}
L&=&\langle \Psi | i\hbar {\partial \over {\partial t}} -H |\Psi \rangle
\nonumber
\\
&=&\int d^3 r \hbar { \dot{\chi} \over 2}\rho_{\chi}+ \cdots
\end{eqnarray}
where $\rho_\chi$ is the number density of electrons participating in the $\chi$ mode, and $\Psi$ is the state vector corresponding to the wave function in Eq.~(\ref{single-valued3}).
This indicates that the conjugate field of $\chi$, $\pi_{\chi}$, is given by
\begin{eqnarray}
\pi_{\chi}={{\delta L} \over {\delta \dot{\chi} }}={ \hbar \over 2} \rho_{\chi}
\end{eqnarray}
Then, the canonical quantization condition 
\begin{eqnarray}
[\hat{\chi}({\bf r},t),\hat{\pi}_{\chi} ({\bf r}',t)]=i\hbar \delta({\bf r}-{\bf r}')
\end{eqnarray}
yields 
\begin{eqnarray}
[\hat{\chi}({\bf r},t),\hat{\rho}_{\chi}({\bf r}',t)]=2i\delta({\bf r}-{\bf r}')
\label{Canonical}
\end{eqnarray}
for the quantization condition for $\hat{\chi}$, 
where $\hat{\rho}_\chi$ is the operator for the number density of electrons that participate in the $\chi$ mode. 
Using the commutation relation in Eq.~(\ref{Canonical}), it can be shown that 
$e^{ {i \over 2}\hat{\chi}}$ is the number changing operators that increases the number of electrons participating the collective motion by one, and $e^{ -{i \over 2}\hat{\chi}}$ decreases the number by one.

By replacing $\theta({\bf r})$ in Eq.~(\ref{field2p}) by $-\hat{\chi}$, the field operators for the new theory are given by
\begin{eqnarray}
\hat{\Psi}_{\uparrow}({\bf r})&=&\sum_{n} e^{-{i \over 2}\hat{\chi} ({\bf r})}\left[ \gamma_{{n} \uparrow } u_{n}({\bf r})  -\gamma^{\dagger}_{{n} \downarrow } v^{\ast}_{n}({\bf r}) \right]
\nonumber
\\
\hat{\Psi}_{\downarrow}({\bf r})&=&
\sum_{n} e^{-{i \over 2}\hat{\chi} ({\bf r})} \left[ \gamma_{{n} \downarrow } u_{n}({\bf r}) +\gamma^{\dagger}_{{n} \uparrow } v^{\ast}_{n}({\bf r}) \right]
\end{eqnarray}
The important point to note is that the above Bogoliubov operators conserve particle numbers;
thus we call them the particle number conserving Bogoliubov operators (PNC-BOs). The PNC-BOs cause the fluctuation of the number of electrons in the collective mode. 

The PNC-BOs satisfy
\begin{eqnarray}
\gamma_{n \sigma}|{\rm Gnd}(N) \rangle=0, \quad  \langle {\rm Gnd}(N)| \gamma^{\dagger}_{n \sigma}=0
\end{eqnarray}
where $N$ is the total number of particles (see Eq.~(\ref{gamma-BCS2}) for the BCS case).
They contain an additional operator $e^{{i \over 2} \hat{\chi}({\bf r})}$.
We take the ground state to be the eigenstate of it that satisfies 
\begin{eqnarray}
e^{{i \over 2} \hat{\chi}({\bf r})}|{\rm Gnd}(N) \rangle= e^{{i \over 2} {\chi}({\bf r})}|{\rm Gnd}(N+1) \rangle
\label{eqBC}
\end{eqnarray}
At this point, we may view the superconducting state in the new theory
as the theory that goes beyond the Schr\"{o}dinger's quantum mechanics by including the quantized version of the neglected $U(1)$ phase by Dirac \cite{DiracSec22}.
As is explained in Ref.~\cite{koizumi2023},
the usual momentum operator adopted by Schr\"{o}dinger can have an emergent gauge field
contribution, which is given as the Berry connection in Eq.~(\ref{Berry}).
The new superconductivity theory includes it in the quantized form given in Eq.~(\ref{Canonical}).

Using the PNC-BO, we can construct the particle number conserving Bogoliubov-de Gennes equations (PNC-BdG equations) as follows: 
Let us consider the following electronic Hamiltonian
\begin{eqnarray}
H_e=\sum_{\sigma} \int d^3 r \hat{\Psi}^{\dagger}_{\sigma}({\bf r}) h({\bf r}) \hat{\Psi}_{\sigma}({\bf r}) 
-{1 \over 2} \sum_{\sigma, \sigma'}\int d^3 r d^3 r' V_{\rm eff}({\bf r}, {\bf r}') \hat{\Psi}^{\dagger}_{\sigma}({\bf r}) \hat{\Psi}^{\dagger}_{\sigma'}({\bf r}') \hat{\Psi}_{\sigma'}({\bf r}') \hat{\Psi}_{\sigma}({\bf r}) 
\nonumber
\\
\end{eqnarray}
where $h({\bf r})$ is the single-particle Hamiltonian given by
\begin{eqnarray}
h({\bf r})={ 1 \over {2m_e}} \left( { \hbar \over i} \nabla +{e \over c} {\bf A} \right)^2+U({\bf r})-\mu 
\end{eqnarray}
and $-V_{\rm eff}$ is the effective interaction between electrons.

We perform the mean field approximation on $H_e$. The result is 
\begin{eqnarray}
H_e^{\rm MF}&=&\sum_{\sigma} \int d^3 r \hat{\Psi}^{\dagger}_{\sigma}({\bf r}) h({\bf r}) \hat{\Psi}_{\sigma}({\bf r}) 
+\int d^3 r d^3 r' 
\left[ \Delta({\bf r}, {\bf r}')\hat{\Psi}^{\dagger}_{\uparrow}({\bf r}) \hat{\Psi}^{\dagger}_{\downarrow}({\bf r}') e^{-{i \over 2}(\hat{\chi}({\bf r}) +\hat{\chi}({\bf r}')) }
+{\rm H. c.} \right]
\nonumber
\\
&+&\int d^3 r d^3 r' 
{ {|\Delta({\bf r}, {\bf r}')|^2} \over {V_{\rm eff}({\bf r}, {\bf r}') }}
\end{eqnarray}
where the gap function $\Delta({\bf r}, {\bf r}')$ is defined as 
\begin{eqnarray}
 \Delta({\bf r}, {\bf r}')= V_{\rm eff}({\bf r}, {\bf r}')\langle e^{{i \over 2}(\hat{\chi}({\bf r}) +\hat{\chi}({\bf r}')) }
\hat{\Psi}_{\uparrow}({\bf r}) \hat{\Psi}_{\downarrow} ({\bf r'}) \rangle
\end{eqnarray}
Due to the factor $ e^{{i \over 2}(\hat{\chi}({\bf r}) +\hat{\chi}({\bf r}'))}$ the expectation value in $\Delta$ can be calculated using the particle number fixed state.

Using commutation relations for $\hat{\Psi}^{\dagger}_{\sigma }({\bf r})$ and $\hat{\Psi}_{\sigma }({\bf r})$,
\begin{eqnarray}
&&\{ \hat{\Psi}_{\sigma }({\bf r}),\hat{\Psi}^{\dagger}_{\sigma' }({\bf r}') \}=\delta_{\sigma \sigma'}\delta({\bf r} -{\bf r}')
\nonumber
\\
&&\{ \hat{\Psi}_{\sigma }({\bf r}),\hat{\Psi}_{\sigma' }({\bf r}') \}=0
\nonumber
\\
&&\{ \hat{\Psi}^{\dagger}_{\sigma }({\bf r}),\hat{\Psi}^{\dagger}_{\sigma' }({\bf r}') \}=0
 \end{eqnarray}
the following relations are obtained
\begin{eqnarray}
\left[\hat{\Psi}_{\uparrow }({\bf r}) , {H}_{\rm MF} \right]&=&
{h}({\bf r})\hat{\Psi}_{\uparrow }({\bf r})+\int d^3 r' \Delta({\bf r},{\bf r}')\hat{\Psi}^{\dagger}_{\downarrow }({\bf r}')e^{-{i \over 2}(\hat{\chi}({\bf r}) +\hat{\chi}({\bf r}')) }
\nonumber
\\
\left[\hat{\Psi}_{\downarrow }({\bf r}) , {H}_{\rm MF} \right] &=&{h}({\bf r})\hat{\Psi}_{\downarrow }({\bf r})-\int d^3 r' \Delta({\bf r},{\bf r}')\hat{\Psi}^{\dagger}_{\uparrow }({\bf r}')e^{-{i \over 2}(\hat{\chi}({\bf r}) +\hat{\chi}({\bf r}')) }
\nonumber
\\
\label{deG1}
\end{eqnarray}

The PNC-BOs $\gamma_{n \sigma}$ and $\gamma^{\dagger}_{n \sigma}$ are the fermion 
operators chosen to satisfy the following relations
\begin{eqnarray}
\left[ {H}_{\rm MF}, \gamma_{n \sigma } \right]=-\epsilon_n \gamma_{n \sigma}, \quad \left[{H}_{\rm MF}, \gamma^{\dagger}_{n \sigma } \right] =\epsilon_n \gamma^{\dagger}_{n \sigma}
\label{deG2}
\end{eqnarray}
Then, ${H}_{\rm MF}$ is diagonalized as
\begin{eqnarray}
{H}_{\rm MF}=E_g + {\sum_{n, \sigma}}' \epsilon_n \gamma^{\dagger}_{n \sigma}\gamma_{n \sigma}
\label{deG3}
\end{eqnarray}
where $E_g$ is the ground state energy, and `$\sum' $' indicates the sum is taken over $\epsilon_n \geq 0$ states.

From Eqs.~(\ref{deG1}), (\ref{deG2}), and (\ref{deG3}), we obtain the following system of equations
\begin{eqnarray}
\epsilon_n u_n({\bf r})&=&
e^{{i \over 2} \hat{\chi}({\bf r})}h({\bf r}) e^{-{i \over 2}\hat{\chi}({\bf r})}u_n({\bf r})+e^{{i \over 2} \hat{\chi}({\bf r})}\int d^3 r' \Delta ({\bf r},{\bf r}')e^{-{i \over 2}\hat{\chi}({\bf r})}v_n({\bf r}')
\nonumber
\\
\epsilon_n v^{\ast}_n({\bf r})&=&-e^{{i \over 2} \hat{\chi}({\bf r})}
 h({\bf r}) e^{-{i \over 2} \hat{\chi}({\bf r})}v^{\ast}_n({\bf r})+e^{{i \over 2} \hat{\chi}({\bf r})}\int d^3r' \Delta ({\bf r},{\bf r}')e^{-{i \over 2}\hat{\chi}({\bf r})}u^{\ast}_n({\bf r}')
 \nonumber
\end{eqnarray}

We take the expectation value of the above equations for the state $|{\rm Gnd}(N) \rangle$. Using the relation in Eq.~(\ref{eqBC}),
the above are cast into the following,
\begin{eqnarray}
\epsilon_n u_n({\bf r})&=&
\bar{h}({\bf r}) u_n({\bf r})+\int d^3 r'\Delta ({\bf r},{\bf r}')v_n({\bf r}')
\nonumber
\\
\epsilon_n v_n({\bf r})&=&-
 \bar{h}^{\ast}({\bf r}) v_n({\bf r})+\int d^3 r'\Delta^{\ast}({\bf r},{\bf r}')u_n({\bf r}')
 \label{e1}
\end{eqnarray}
where the single particle Hamiltonian $\bar{h}$ is
\begin{eqnarray}
\bar{h}({\bf r})={ 1 \over {2m_e}} \left( { \hbar \over i} \nabla +{e \over c} {\bf A}-{ \hbar \over 2} \nabla \chi \right)^2+U({\bf r})-\mu 
 \label{e2}
\end{eqnarray}
the pair potential $\Delta({\bf r}, {\bf r}')$ is
\begin{eqnarray}
\Delta({\bf r}, {\bf r}')=V_{\rm eff}{\sum_n}' \left[ u_n({\bf r}) v^{\ast}_n({\bf r}')(1- f(\epsilon_n))-u_n({\bf r}') v^{\ast}_n({\bf r})f(\epsilon_n) \right]
 \label{e3}
\end{eqnarray}
and $f(\epsilon_n)$ is the Fermi function ($T \rightarrow 0$ limit should be considered for the ground state). The number density is given by
\begin{eqnarray}
\rho({\bf r})={\sum_n}' \left[ |u_n({\bf r})|^2f(\epsilon_n)+|v_n({\bf r})|^2({\bf r})(1- f(\epsilon_n)) \right]
 \label{e4}
\end{eqnarray}
 They are the PNC-BdG equations \cite{koizumi2019}.
The gauge potential in the single particle Hamiltonian $\bar{h}({\bf r})$ is the effective one given by
\begin{eqnarray}
{\bf A}^{\rm eff}={\bf A}- { {\hbar } \over {2e}} \nabla \chi
\end{eqnarray}
 This combination corresponds to Eq.~(\ref{theta-Nambu}), and appeared in the velocity field in Eq.~(\ref{eq12v2}).

Now the total energy is a functional of ${\bf A}- { {\hbar } \over {2e}} \nabla \chi$,
its time-component partner $\varphi + { {\hbar } \over {2e}} \partial_t \chi$, and $\rho$.
We may write it in the following form
\begin{eqnarray}
H=\int d^3 r \left[ {\cal H} \left( {\bf A}- { {\hbar } \over {2e}} \nabla \chi, \varphi + { {\hbar } \over {2e}} \partial_t \chi, \rho \right)-e\rho \varphi \right]
\end{eqnarray}
Note that the last term is the contribution from the electrostatic potential, and the number density $\rho$ is the sum of $\rho_\chi$ and the rest.
The above functional should be compared with the one proposed by Oliveira, Gross, and Kohn \cite{KS-BdG}, which uses $\Delta$ as the extra parameter to describe the superconducting state.
In the present case, the superconducting state is characterized by $(\nabla \chi, \partial_t \chi)$; they originate from the neglected $U(1)$ phase by Dirac \cite{koizumi2023}, which upsets
the fundamental assumption of the density functional theory \cite{Hohenberg-Kohn} since the use of the operator form of the momentum ${\bf p}=-i\hbar \nabla$ is violated. 
In other words, the present formalism indicates that the superconducting state is related to the violation of using ${\bf p}=-i\hbar \nabla$. 
Note that the present formalism agrees with the claim that the fundamental physics occurring in superconductivity
is the gauge symmetry breaking accompanied by the appearance of the Nambu-Goldstone mode \cite{Anderson,Weinberg}.

Employing the classical counterpart of the commutation relation in Eq.~(\ref{Canonical}), i.e.,
taking ${\hbar \over 2} {\chi}$ and $ \rho_\chi$ as canonical conjugate classical variables, we obtain the following Hamilton equations
\begin{eqnarray}
{\hbar \over 2} \partial_t {\chi}&=&{{\partial {\cal H} } \over {\partial \rho_\chi}}-e\varphi
\label{H1}
\\
 \partial_t {\rho_\chi}&=&\nabla \cdot { {\partial {\cal H}} \over {\partial {{\hbar \over 2}\nabla\chi}}}=-{1 \over e}\nabla \cdot { {\partial {\cal H}} \over {\partial {\bf A}}}
 \nonumber
 \\
 &=&{1 \over e}\nabla \cdot {\bf j}
\label{H2}
\end{eqnarray}
The equation (\ref{H1}) corresponds to Eq.~(\ref{eq1mu}) with $\mu={{\partial {\cal H}} \over {\partial \rho_\chi}}$.
The equation (\ref{H2}) describes the local charge conservation with ${\bf j}=-e\rho_\chi{\bf v}$, where
${\bf v}$ is given by Eq.~(\ref{eq12v2}).

\end{appendices}



\end{document}